\documentclass[aps,prd,amsmath,amssymb,10pt,letterpaper,balancelastpage,showpacs,superscriptaddress,nofootinbib,floatfix,notitlepage,preprintnumbers,longbibliography,twocolumn]{revtex4-2}

\usepackage{graphicx}
\usepackage{soul}
\usepackage{color}
\usepackage{xcolor}
\usepackage{enumitem}
\usepackage{cancel}
\usepackage{mathtools}
\usepackage{hyperref}
\usepackage{orcidlink}
\usepackage{cases}
\usepackage[normalem]{ulem}
\usepackage{bbm}

\usepackage[T1]{fontenc}

\usepackage{tabularray}
\usepackage{multirow}

\usepackage{makecell}

\usepackage[caption=false]{subfig}

\usepackage[export]{adjustbox}

\usepackage{tabularray}
\UseTblrLibrary{booktabs}

\newcommand{\pfrac}[2]{{\left(\frac{#1}{#2}\right)}}

\newcommand{\bra}[1]{\left\langle #1 \right|}
\newcommand{\ket}[1]{\left|#1\right\rangle}

\newcommand{\expect}[1]{\left\langle#1\right\rangle}

\newcommand{\nk}{\hat{n}_k}
\newcommand{\bk}{\mathbf{k}}
\newcommand{\ramp}{{\rm ramp}}
\newcommand{\free}{{\rm free}}

\newcommand{\sigE}{\mathcal{E}}
\newcommand{\sigErms}{\mathcal{E}_{\rm rms}}
\newcommand{\sigEmin}{\mathcal{E}_{\rm min}}
\newcommand{\sigEmax}{\mathcal{E}_{\rm max}}
\newcommand{\DD}{{\rm DD}}
\newcommand{\Nbins}{N_{\rm bins}}
\newcommand{\omof}{\omega_{\rm offset}}
\newcommand{\dchar}{\delta^*}
\newcommand{\QSP}{{\rm QSP}}

\begin{document}

\preprint{FERMILAB-PUB-26-0815-SQMS-T}

\title{A Grover-Accelerated Search for Dark Matter}
\date{\today}

\author{Zachary Bogorad\,\orcidlink{0000-0001-9913-6474}\,}
\email{zbogorad@fnal.gov}
\affiliation{Fermilab Theory Division, Fermi National Accelerator Laboratory, Batavia, IL 60510, USA}
\affiliation{Superconducting Quantum Materials and Systems Center (SQMS), Fermi National Accelerator Laboratory, Batavia, IL 60510, USA}

\author{Roni Harnik\,\orcidlink{0000-0001-7293-7175}\,}
\email{roni@fnal.gov}
\affiliation{Fermilab Theory Division, Fermi National Accelerator Laboratory, Batavia, IL 60510, USA}
\affiliation{Superconducting Quantum Materials and Systems Center (SQMS), Fermi National Accelerator Laboratory, Batavia, IL 60510, USA}

\begin{abstract}
\noindent We show that Grover's algorithm can be used to parametrically accelerate wave-like dark matter searches. Despite having a finite coherence time~$\tau$, wavelike dark matter can imprint a coherent~$O(1)$ phase onto a sensor through a dispersive coupling over an integration time which significantly exceeds~$\tau$. As such, we show that dark matter can be used as a repeatable quantum phase gate in a quantum device. If the sensor is prepared in a superposition of frequencies, we show that this phase can be transferred to a qubit register and digitized. The qubits are thus available for quantum computational processing in which dark matter can take part as a coherent gate. We show that Grover's algorithm can be used to speed up a search for dark matter in~$\Nbins$ frequency bins by a factor of~$\sqrt{\Nbins}$. Although the technological challenges to implementing such a search remain substantial, this demonstrates that there is no fundamental obstruction to accelerating wave-like dark matter searches with quantum computational techniques over many coherence times.
\end{abstract}

\maketitle

\section{Introduction}\label{sec:Introduction}

\begin{figure}[t]
    \centering
    \includegraphics[width=0.99\linewidth]{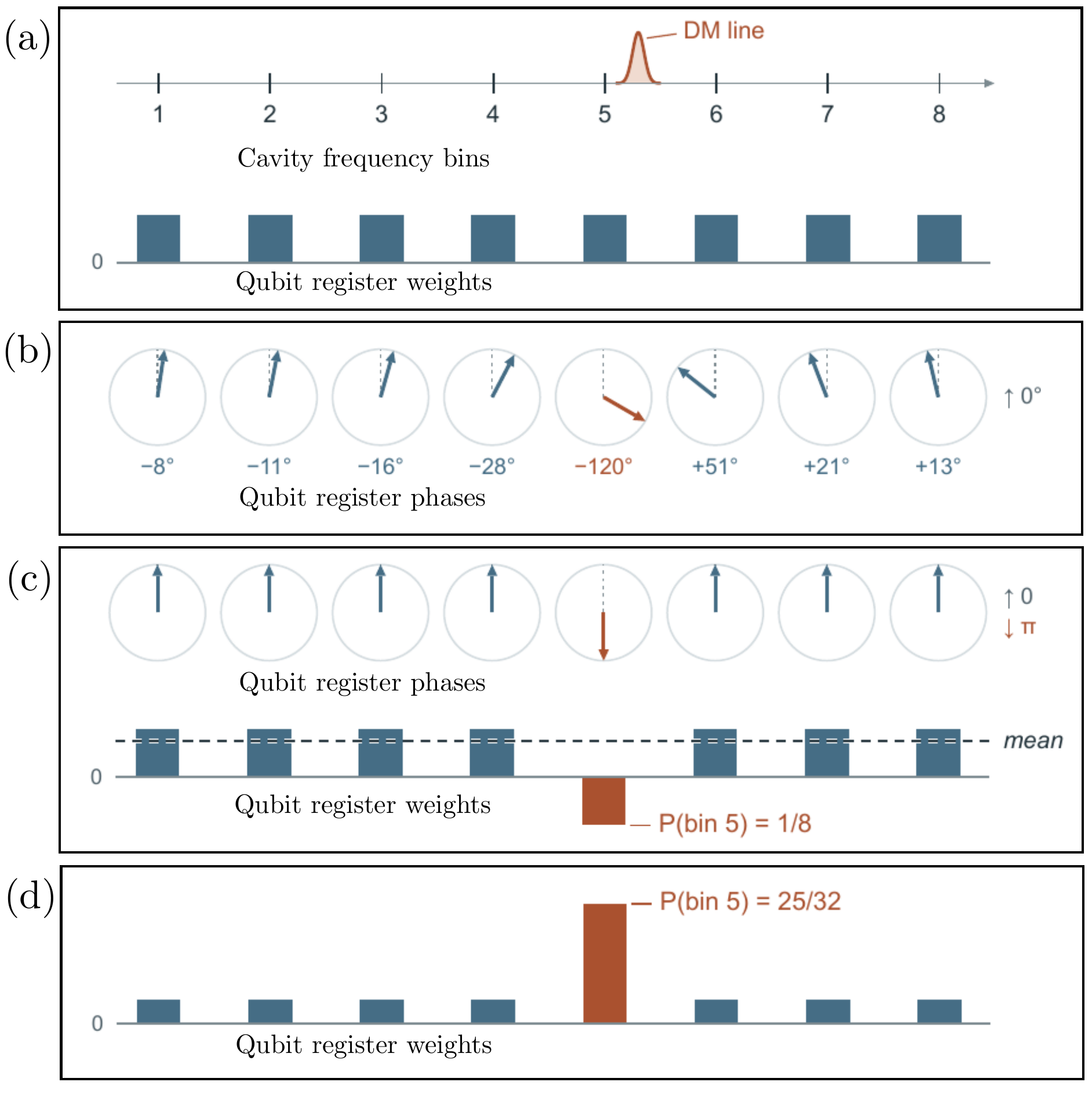}
    \caption{An outline of a Grover-based dark matter search. (a)~A cavity haloscope is prepared in a vacuum state with a superposition of frequencies by entangling it with a qubit register. Eight frequency bins and their corresponding register entries are shown. (b)~The cavity accumulates a dark matter-induced phase over many DM coherence times and then returns to the vacuum state, leaving the phases imprinted in the qubit register via its dispersive coupling. The bin closest to the dark matter mass registers the largest phase. (c)~The qubit phases are digitized to 0 or $\pi$ using quantum signal processing (QSP). (d)~The signal bin, with phase $\pi$, is amplified via Grover's algorithm.}
    \label{fig:placeholder}
\end{figure}

The existence of dark matter is well established, but its microscopic nature remains unknown. Well-motivated candidates include wavelike dark matter candidates, such as axions, which arise in a solution to the strong CP problem~\cite{peccei-quinn}, and dark photons, which can interact with ordinary matter through a kinetic mixing~\cite{holdom-dark-photon}. Both of these examples have compelling cosmological production mechanisms~\cite{Preskill1983,AbbottSikivie1983,DineFischler1983, NelsonScholtz2011}. At sufficiently small masses, their large occupation numbers allow dark matter to be described as a classical field.

Axions and dark photons can produce weak, oscillatory signals in laboratory detectors. Resonant experiments, such as cavity haloscopes~\cite{Sikivie1983}, exploit this behavior to achieve high sensitivity within a narrow frequency range, but the unknown mass of dark matter, and thus its unknown oscillation frequency, requires typical experiments to perform a large number of sequential measurements with different resonant frequencies; the time required to cover the range of possible dark matter masses is therefore a central limitation of these searches.

Quantum techniques have already improved dark matter searches through squeezed-state receivers and superconducting-qubit photon counting~\cite{Backes2021,Dixit2021,Braggio2025}. Quantum computation additionally offers the ability to reduce this repetition count, and thus to speed up a wave-like dark matter search: Ref.~\cite{Allen:2025hdx} recently introduced a sensing protocol that converts an interaction with an unknown-frequency oscillatory field into an operation suitable for Grover's algorithm~\cite{GroverOriginal}. By coherently processing the phases acquired in a detector placed in a superposition of resonant frequencies, this approach achieves a parametric advantage over sequential scanning, establishing the Grover--Heisenberg limit for their sensing problem. Related protocols have since been demonstrated in superconducting qubit--cavity systems~\cite{Xu2026,Ouellet2026}. Unfortunately, the exact algorithm in Ref.~\cite{Allen:2025hdx} is not immediately applicable to dark matter, as it assumes an infinite (or at least extremely long) coherence time for the signal. Galactic dark matter has a velocity-induced frequency spread, and therefore its phase changes on a timescale $\tau \sim 10^6/\mu$ for a dark matter mass (and thus frequency) $\mu$~\cite{Gramolin2022}. For the weak signals that dark matter could induce, a meaningful detector response requires integration over many coherence times. It is therefore necessary to determine whether this response can remain available for coherent quantum processing despite the changing phase of the incident field.

In this work, we show that Grover's algorithm can be used to speed up dark matter searches in spite of this finite coherence time. We do so using the example of a dispersive interaction of qubits with a cavity, although the same principles generalize straightforwardly to other sensing platforms. Dark matter drives a displacement of the cavity state whose direction changes with its phase, and also produces an energy shift proportional to the field intensity and inversely proportional to the difference between the dark matter mass and the cavity's resonant frequency. This energy shift causes a relative phase to accumulate in a coupled qubit register, which can accumulate over many dark matter coherence times. By adiabatically controlling the effective interaction between dark matter and the cavity, we can ultimately return the cavity to vacuum while retaining the accumulated register phase. The resulting operation preserves superpositions of qubit states corresponding to different resonant frequencies of the cavity, and can be incorporated into subsequent quantum computation.

Our construction is based on a simplified cavity-haloscope model whose frequency is controlled by a register of dispersively coupled qubits. A photon blockade~\cite{Chakram:2020eky} or Kerr nonlinearity~\cite{Kirchmair:2012rsm} restrict the cavity Hilbert space to its first two levels, modifying the effect of dark matter on the cavity. We then use quantum signal processing~\cite{Motlagh:2023oqc,Martyn:2021eaf,Haah:2019pes} to convert the accumulated phases into an approximate binary marking of signal-containing frequency bins, enabling a Grover search. Subject to the coherence and control requirements described below, the protocol reduces the search time over $\Nbins$ frequency bins by a factor scaling as $\sqrt{\Nbins}$, up to constants and logarithmic overheads, relative to the standard sequential strategy of observing one frequency at a time. Maintaining the register's coherence throughout the search presents enormous technical challenges for the specific detector model considered here, and modifications of our protocol will be necessary in order to experimentally realize a Grover-accelerated dark matter search. Nonetheless our results establish a proof of principle, showing that the finite coherence time of dark matter does not itself preclude a quantum computational enhancement of the search.

The remainder of this paper is organized as follows. Section~\ref{sec:Detector} introduces our toy detector model, which we use for concreteness throughout the rest of the work. Section~\ref{sec:CoherentPhaseAcc} demonstrates how this model allows one to coherently accumulate phases for times much longer than the dark matter coherence time $\tau$. Section~\ref{sec:GroverSearch} then explains how this coherent phase accumulation can be built into a complete Grover-based search procedure, including follow-up measurements needed to verify signals and to adapt to unknown signal strengths. Section~\ref{sec:TechReqs} briefly examines the principal sources of error in such a measurement, and discusses the resulting technological requirements. Finally, we conclude in Section~\ref{sec:Conclusion}. We use natural units $\hbar = c = k_B = 1$ everywhere in the text.

\section{Simplified Detector Model}\label{sec:Detector}

Our toy detector model is based on a cavity haloscope (although the same ideas can likely be applied to many other detection platforms); in essence, a high-quality electromagnetic cavity that couples to dark photons, or, if one applies an additional magnetic field across it, to axions. Cavity haloscopes are a widely-used detector class for both dark photons and axions due to their large volume and high quality; see for example Refs.~\cite{Sikivie1983,Du2018,Brubaker2017,Backes2021,Rettaroli2024,Cervantes2024,Dixit2021,Tang2024}. The effect of dark matter---whether that is a dark photon or an axion---is to apply a not-quite monochromatic, time-dependent displacement operator to the cavity, as explained in, for example, Refs.~\cite{Dixit2021,Freiman2026,BogoradHarnik2026,MaSheng2026}; we also review this below.  For simplicity, most of our discussion will assume the cavity is at zero temperature; we discuss finite-temperature effects in Section~\ref{sub:CavityErrors}.

\begin{figure}
    \centering
    \includegraphics[width=0.99\linewidth]{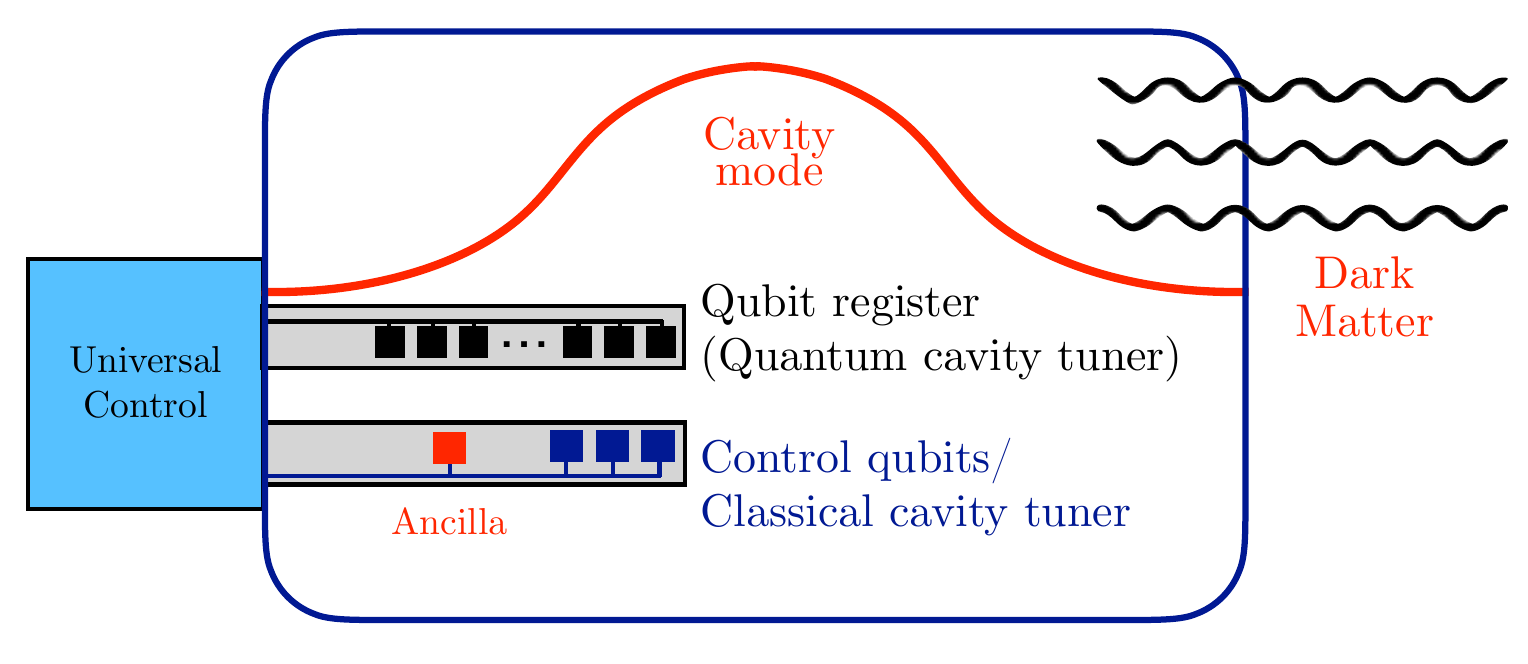}
    \caption{A sketch of the simplified detector model. A high quality cavity mode interacts with dark matter linearly. The cavity frequency is controlled by a qubit register which enables a superposition of cavity frequencies and will act as the Grover register. Additional qubits are used to execute frequency bin offsets and to dynamically ramp the signal on and off to achieve adiabatic evolution.}
    \label{fig:DeviceSketch}
\end{figure}

As in many haloscope-based dark matter searches, we will employ only a single mode of the cavity. Instead, we will engineer our system so that one mode can be in a superposition of frequencies. This superposition will be arranged via a collection of $n_q + {\rm (a~few)}$ dispersively-coupled qubits (e.g. transmons~\cite{Koch2007,Blais2004}), which act in effect as a qubit state-dependent frequency shift for the cavity. Classically controlled electronic tuning of cavity resonances has already been demonstrated in dark matter searches, using flux-controlled superconducting circuits~\cite{Zhao2025,Nakazono2025} and off-resonant sideband drives~\cite{Zheng2026}.

The first $n_q$ of these qubits, which we will refer to as the ``register'' and label as number $0$ through $n_q-1$, will create the comb of effective cavity frequencies over which we will search. To achieve this we arrange the dispersive couplings of these qubits in powers of $2$, such that we can conveniently interpret the frequency shift as a binary number set by the qubits' states, as we show below.

The last few ``control'' qubits will be used as classical cavity tuners and to implement cavity operations; we will generally not be interested in their dynamics, but they are otherwise analogous to the register qubits. Two will be used for an overall frequency shift (for reasons discussed in Section~\ref{sub:Binning}), one will be used to effectively modulate the strength of dark matter's coupling to the cavity and allow an adiabatic on/off ramp of the signal, as explained in Section~\ref{sub:Ramping}, and a few others (depending on implementation details beyond the scope of this paper) are used throughout our work in order to mediate changes in the cavity state. We will also later need one ``ancilla'' qubit which we single out since its dynamics will be important to us.

The combined Hamiltonian for the cavity and the register qubits (we omit the control qubits, since we will not be interested in their dynamics, as well as the ancilla, which we introduce later) is then
\begin{align}
    H &= H_{\rm cav} + H_{\rm qubit} + H_{\rm disp} + H_{\rm DM} \label{eq:Htotal}
\end{align}
where
\begin{subequations}
\begin{align}
    H_{\rm cav} &= \left( \omega_c + \omof \right) \hat{a}^\dagger \hat{a} \label{eq:Hcav} \\
    H_{\rm qubit} &= \omega_q \sum_{k=0}^{n_q-1} \ket{1_k}\bra{1_k} =: \omega_q \sum_{k=0}^{n_q-1} \nk \label{eq:Hqubit} \\
    H_{\rm disp} &= \sum_{k=0}^{n_q-1} \chi_k \nk \hat{a}^\dagger \hat{a} = \chi_0 \sum_{k=0}^{n_q-1} 2^k \nk \hat{a}^\dagger \hat{a} \label{eq:Hdisp} \\
    H_{\rm DM} &= 2\epsilon(t) \cos\left(\mu t + \phi(t)\right) \left(\hat{a}^\dagger + \hat{a}\right) \label{eq:HDM}
\end{align}
\end{subequations}
where $\ket{1_k}$ is one of the two states of the $k$'th qubit, we have defined $\hat{n}_k = \ket{1_k}\bra{1_k}$ for convenience, and $\chi_k = 2^k\chi_0$ is the dispersive coupling of the $k$'th qubit to the cavity. The effects of the control qubits are absorbed into the frequency shift $\omof$, since we are not interested in their dynamics. 

$\epsilon(t)$ and $\phi(t)$ parametrize the dark matter-cavity interaction, which is approximately oscillatory at a frequency $\mu$, but has a finite coherence time $\tau$ of order $10^6 / \mu$, corresponding to a slowly-drifting phase $\phi(t)$. $\mu$ is approximately, but not exactly, the dark matter mass, although we will sometimes refer to it as a mass for brevity; more precisely, a dark matter component with speed $v_{\rm DM}$ oscillates at angular frequency $m_{\rm DM}(1+v_{\rm DM}^2/2)$, so the signal is broadened and shifted above the rest mass by a fraction of order $10^{-6}$. We use $\mu$ as a representative frequency within this line, with the frequency spread around it encoded in $\epsilon(t)$ and $\phi(t)$.

To connect $\epsilon(t)$ to more familiar properties of dark matter, recall that, for a vacuum cavity of volume $V$, the signal amplitude is~\cite{Du2018, Cervantes2024}
\begin{align}
    \overline{|\epsilon(t)|^2} \approx \frac{\rho_{\rm DM}V}{4\mu}
    \begin{cases}
        g_{a\gamma\gamma}^2 B_0^2 C_a,
        & \text{axions},\\[2pt]
        \kappa^2\mu^2 C_{A'},
        & \text{dark photons},
    \end{cases}
\end{align}
where the overline represents a time average, $\rho_{\rm DM}$ is the local dark matter density, $g_{a\gamma\gamma}$ is the axion–photon coupling, $B_0$ is the magnitude of the applied, uniform magnetic field, and $\kappa$ is the dark photon kinetic mixing. The amplitude-squared of the dark matter field is set by $\rho_\mathrm{DM}$ on long time scales, but experiences order one modulations on the same timescale $\tau$~\cite{Foster2021Interferometry,Foster2018Halo,Derevianko2018Correlations,Cheong2025WaveDM}. In our scheme, we will consider integration times much longer than $\tau$ and these variations will average down.  The dimensionless mode-overlap factors are
\begin{subequations}
\begin{align}
    C_a &= \frac{ \left| \int dV \,\mathbf E_c \cdot \hat{\mathbf B}_0\right|^2}{V \int dV \, \left|\mathbf E_c\right|^2 } \\[8pt]
    C_{A'} &=
    \frac{ \overline{ \left| \int dV \, \mathbf E_c \cdot \hat{\mathbf e}_{A'} \right|^2} }{V \int dV \, \left| \mathbf E_c \right|^2},
\end{align}
\end{subequations}
with $\mathbf E_c$ the cavity-mode electric-field profile, $\hat{\mathbf B}_0$ the direction of the applied magnetic field, and $\hat{\mathbf e}_{A'}$ the dark photon polarization; the average in $C_{A'}$ is needed to handle time-varying dark photon polarization~\cite{Caputo2021}. As is the case for the field amplitude, order one variations in dark photon couplings are averaged down for the long exposures we will consider. 

Later in this work, we will need one additional contribution to our Hamiltonian, in order to effectively remove the second excited Fock level $\ket{2_c}$ of the cavity, essentially turning it into a qubit. This can, in principle, be done in at least two ways: by implementing a photon blockade that effectively splits the $\ket{2_c}$ Fock level~\cite{Chakram:2020eky}, or by adding a Kerr nonlinearity $H_{\rm Kerr} \propto \hat{a}^\dagger \hat{a} (\hat{a}^\dagger \hat{a}-1)$ that shifts the energy of all levels with $\hat{a}^\dagger \hat{a} > 1$~\cite{Kirchmair:2012rsm}. We defer discussion of this term until we need it in Section~\ref{sec:GroverSearch}, and feasibility discussion to Section~\ref{sec:TechReqs}.

We will often want to discuss the behavior of the component of the cavity wavefunction entangled with a particular set of qubits in $\ket{1}$, so we define
\begin{align}
    \bk = \sum_{\hat{n}_k = 1} 2^k.
\end{align}
We thus describe the state of all $n_q$ qubits with a single binary number with $n_q$ digits, which we call $\bk$. We will use these states of definite $\hat{n}_k$ for all~$k$ as a convenient basis; each of the $2^{n_q}$ basis states is defined by a single~$\bk$. Conditional on the qubits being in $\ket{\bk}$, the cavity then has an effective frequency
\begin{align}
    \omega_\bk = \omega_c + \omof + \chi_0\bk.
\end{align}

We now go into a corotating frame with a frequency $\omega_\bk$ in order to simplify the interaction. First, note that in the interaction picture of $\omega_\bk \hat{a}^\dagger \hat{a}$ we can replace
\begin{align}
    H \supset \omega_\bk \hat{a}^\dagger \hat{a} + 2\epsilon(t) \cos\left(\mu t + \phi(t)\right) \left(\hat{a}^\dagger + \hat{a}\right)
\end{align}
with
\begin{align}
\begin{split}
    H_I &\supset \epsilon(t) \left( e^{-i(\mu-\omega_\bk)t-i\phi(t)} \hat a^\dagger + e^{i(\mu-\omega_\bk)t+i\phi(t)} \hat a \right) \\
    &\hphantom{\supset} + \epsilon(t) \left( e^{i(\mu+\omega_\bk)t+i\phi(t)} \hat a^\dagger + e^{-i(\mu+\omega_\bk)t-i\phi(t)} \hat a \right).
\end{split}
\end{align}
The terms in the lower row oscillate at a frequency
$\omega_\bk + \mu$, which for our purposes will be very large, i.e.
\begin{align}
    \epsilon(t), \, |\omega_\bk-\mu|, \, \frac{1}{\tau} \ll \omega_\bk + \mu.
\end{align}
They will therefore rapidly average away and we will omit them from this point on (although we consider deviations from this approximation in Section~\ref{sub:CounterRotatingTerms}). Conversely, the terms in the upper row oscillate at the frequency
\begin{align}
    \delta_\bk := \omega_\bk - \mu,
\end{align}
which may be small enough to be relevant; these are the terms we will keep. We will call $\delta_\bk$ the detuning of the~$\bk$'th cavity frequency. If we now perform an additional frame rotation by $\delta_\bk$, going into the frame co-rotating at frequency $\mu$, this leaves us with
\begin{align}
    H_{\rm eff} = \delta_\bk \hat{a}^\dagger \hat{a} + \sigE(t) \hat{a}^\dagger  + \sigE(t)^* \hat{a} \label{eq:effHcav}
\end{align}
where
\begin{align}
    \sigE(t) := \epsilon(t) e^{-i\phi(t)}.
\end{align}
This is the effective Hamiltonian of the qubit-conditioned cavity mode coupled to dark matter, and will be the starting point for our subsequent discussion.

Note, however, that our initial state will not be one of definite $\bk$, but rather a state that is an equal superposition of every $\bk$:
\begin{align}
    \bigotimes\limits_{k=0}^{n_q-1} \ket{+_k} = \bigotimes\limits_{k=0}^{n_q-1} \pfrac{\ket{0_k} + \ket{1_k}}{\sqrt{2}} &= \frac{1}{\sqrt{2^{n_q}}}\sum_{\bk = 0}^{2^{n_q}-1} \ket{\bk}.
\end{align}
This will create the frequency comb we desire, in which some component of the cavity wavefunction---specifically the one entangled with $\ket{\bk}$---will simultaneously have an effective frequency $\omega_\bk$ (in the laboratory frame), or an effective detuning $\delta_\bk$ in the co-rotating frame.

\section{Coherent Phase Accumulation Beyond Dark Matter's Coherence Time}\label{sec:CoherentPhaseAcc}

We now describe how our toy model of a qubit-coupled haloscope can be used to coherently accumulate a dark matter-induced phase for much longer than dark matter's coherence time. This is a fundamentally different sensing strategy from that used in existing dark matter experiments, which do not preserve coherence beyond the dark matter coherence time; we elaborate on this difference in Appendix~\ref{app:CavityEvolution}. Later in this work, we will use this novel strategy (with a small modification) in order to implement Grover's algorithm to search for dark matter. However, we emphasize that this is merely an example, and it may be possible to combine this technique with other inherently quantum mechanical techniques as well. We leave exploring this to future work.

Note that we do not need the photon blockade or Kerr nonlinearity in this section; those will only become necessary for implementing Grover's algorithm specifically.

\subsection{Free Dispersive Evolution}\label{sub:FreeEvolution}

\begin{figure}[t]
    \centering
    \includegraphics[width=\linewidth]{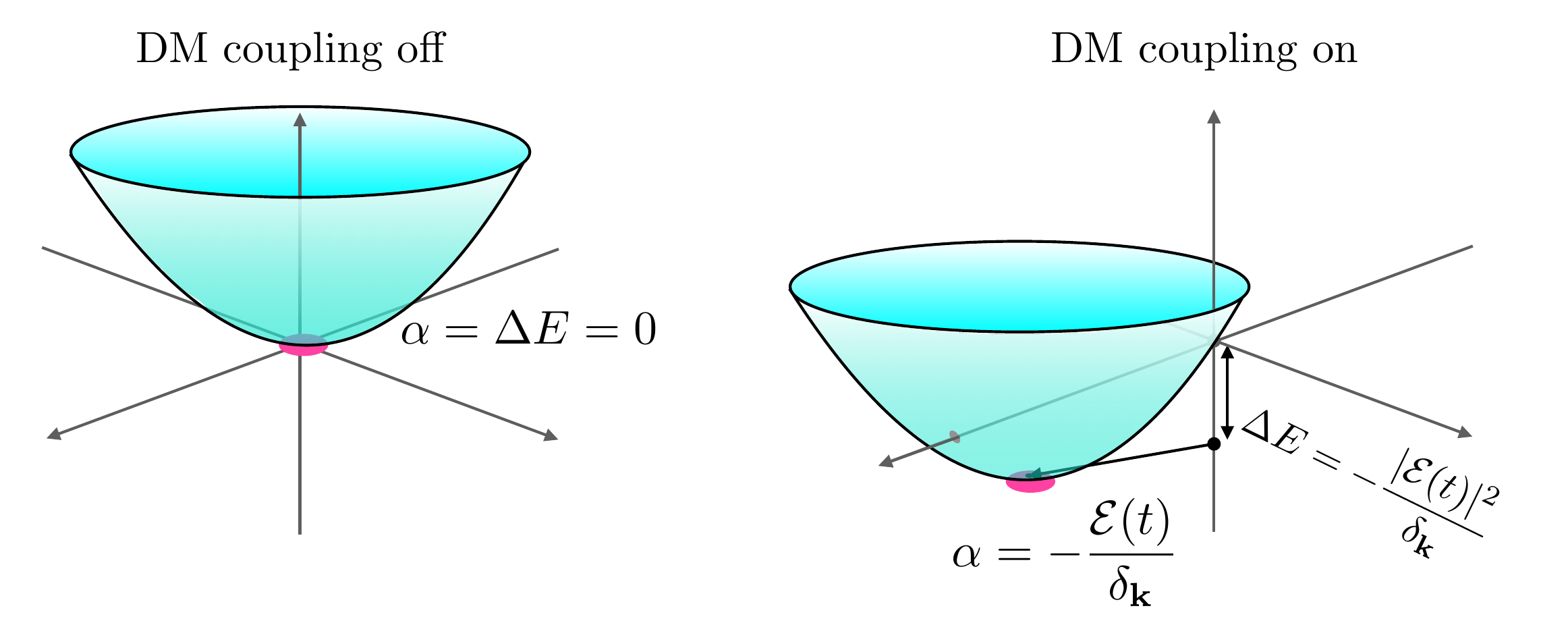}
    \caption{Plots of the cavity Hamiltonian as a function of the $x$ and $p$ quadratures in the frame rotating at the frequency of the dark matter mass, for a positive detuning $\delta_\bk>0$. In the absence of a dark matter signal (left), the minimum is a vacuum state. In the presence of dark matter (right), the minimum is shifted in a time-dependent direction in the $x$-$p$ plane, but the energy is shifted by a constant $|\sigE|^2/\delta_\bk$. Our protocol adiabatically transitions between these two.}
    \label{fig:shift}
\end{figure}

It is now helpful to rewrite Eq.~\eqref{eq:effHcav} by completing the square:
\begin{align}
    H_{\rm eff} = \delta_\bk \left( \hat{a} + \frac{\sigE(t)}{\delta_\bk} \right)^\dagger \left( \hat{a} + \frac{\sigE(t)}{\delta_\bk} \right) - \frac{|\sigE(t)|^2}{\delta_\bk}.
\end{align}
As shown in Figure~\ref{fig:shift}, the instantaneous effect of dark matter is thus twofold: (1) to shift the cavity's lowest- energy state\footnote{Note that for negative detuning, $\delta_\bk < 0$, this is the highest-energy state, not the lowest, but in fact both statements hold only within the co-rotating approximation. For $\delta_\bk > 0$, cavity decays correspond to lowering the energy by the full $\omega_\bk \gg \delta_\bk$, and thus decays from this state are still possible. We discuss these in Section~\ref{sub:CavityErrors}. For $\delta_\bk < 0$, it is likewise the case that sufficiently occupied states will have higher energies.} to a coherent state with $\alpha = -\sigE(t)/\delta_\bk$, and~(2) a corresponding energy shift of $-|\sigE(t)|^2/\delta_\bk$. Note that the directional displacement of the minimum with respect to the origin depends on the dark matter phase $\phi(t)$, which changes smoothly by $O(1)$ every dark matter coherence time $\tau_\mathrm{DM}$. By contrast, the energy shift is independent of the dark matter phase and remains constant so long as the effective  dark matter coupling and frequency are constant~\cite{MasiaRoig2023}. This allows us to read out a coherent signal from dark matter over many dark matter coherence times. 

In order to take advantage of this coherent energy shift, we will adiabatically follow this local extremum for a time $T_\free$,  evolving each of the qubit-conditioned states as
\begin{align}
    \ket{\bk}\otimes\ket{\alpha(t=0)} \longrightarrow e^{i\Phi_\bk} \ket{\bk}\otimes\ket{\alpha(t=T_\free)}\label{eq:phase}
\end{align}
where $\alpha(t)$ is the position of the local extremum at time $t$ and
\begin{align}
    \Phi_\bk = \int dt \, \frac{|\sigE(t)|^2}{\delta_\bk} = \frac{\sigErms^2 T_\free}{\delta_\bk}, \label{eq:FreeEvolPhase}
\end{align}
with $\sigErms$ the root-mean-square magnitude of $\sigE(t)$~\cite{Pechal2012}. (We assume for now that it is safe to take this average; as we discuss in Section~\ref{sec:TechReqs}, this is a safe assumption for both axions and dark photons.) We will therefore set
\begin{align}
    T_\free \sim \frac{\dchar}{\sigErms^2}
\end{align}
in order to acquire an order-one (more precisely, order-$\pi$, but we use these terms interchangeably) phase at a characteristic detuning $\dchar$, which we define in Section~\ref{sub:Binning}. 
In order to construct a ``clean'' Grover oracle, we will then want to ``digitize'' this phase---transforming all small phases to 0 and any large phase to $\pi$---which we do in Sections~\ref{sub:QSPGeneral} and~\ref{sub:QSPforDM}.

We first need to ensure that the cavity will, in fact, adiabatically follow this instantaneous extremum during the free evolution. Since the co-rotating effect of the dark matter is a pure displacement (see Eq.~\eqref{eq:HDM}), a cavity that begins in a coherent state will remain in one exactly (up to decays, thermal excitation, etc.; see Section~\ref{sec:TechReqs}). We therefore need only to evaluate $\alpha(t) = \expect{a}_t$, which is straightforward:
\begin{align}
    \frac{d\alpha}{dt} = i\expect{\left[H_{\rm eff},a\right]} =  -i\delta_\bk \alpha - i\sigE, \label{eq:dalphadt}
\end{align}
where it is intuitively useful to think of the first term on the right-hand-side as a rotation of the plane around the origin, and the second term as a translation along the direction set by $\sigE$. For any choice of rotation and translation rate there will be a point on the plane in which the two cancel, which will be a stationary point. Our goal will be to track this point adiabatically.

It is helpful to rewrite this in terms of the deviation of $\alpha$ from the instantaneous stationary point, $\Delta\alpha(t) = \alpha(t) + \sigE(t)/\delta_\bk$. Then
\begin{align}
    \frac{d\Delta\alpha}{dt} = -i\delta_\bk \Delta\alpha + \frac{1}{\delta_\bk}\frac{d\sigE}{dt}. 
\end{align}
The magnitude of $d\sigE/dt$ is set simply by dark matter's coherence time:
\begin{align}
    \left| \frac{d\sigE}{dt} \right| \sim \frac{\sigErms}{\tau},
\end{align}
which is simply the statement that $\sigE(t)$ changes by order one on the timescale $\tau$. We will work in the regime when $|\delta_\bk| \tau \gg 1$, i.e. where $\sigE$ changes slowly compared to the detuning frequency. Then
\begin{align}
    \Delta\alpha \sim -\frac{i\sigE}{\tau \delta_\bk^2}
\end{align}
corresponding to a fractional error
\begin{align}
    \frac{|\Delta\alpha|}{\left|\sigE/\delta_\bk\right|} \sim \frac{1/\tau}{|\delta_\bk|}.   
\end{align}
Thus as long as we remain in the regime\footnote{In fact it is likely sufficient to take $\delta_\bk\tau$ of order a few or even one, but we leave precise numerical evaluation of this borderline regime to future work. We will treat $\delta_\bk\tau$ as an order-one number for our sensitivity estimates.}
\begin{align}
    |\delta_\bk| \tau \gg 1, \label{eq:AdiabaticityDetuningCondition}
\end{align}
we can safely assume that the cavity adiabatically follows the instantaneous stationary point and acquires the phase $\Phi_\bk$ set by Eq.~\eqref{eq:FreeEvolPhase}. Note that this condition is parametrically stronger than that in Ref.~\cite{Allen:2025hdx}, which requires only $\sigErms/|\delta_\bk| \ll 1$; this is the key change that allows the protocol to accommodate a finite signal coherence time. Although we will not necessarily always be in this regime---since the dark matter frequency is not known, it is always possible that it happens to have a tiny detuning $\delta_\bk \ll 1/\tau$ for some $\bk$---we show in Section~\ref{sub:Binning} that we can operate the detector in a way that renders its behavior for $\delta_\bk \ll 1/\tau$ irrelevant.

\begin{figure*}
    \centering
    \includegraphics[width=0.95\linewidth]{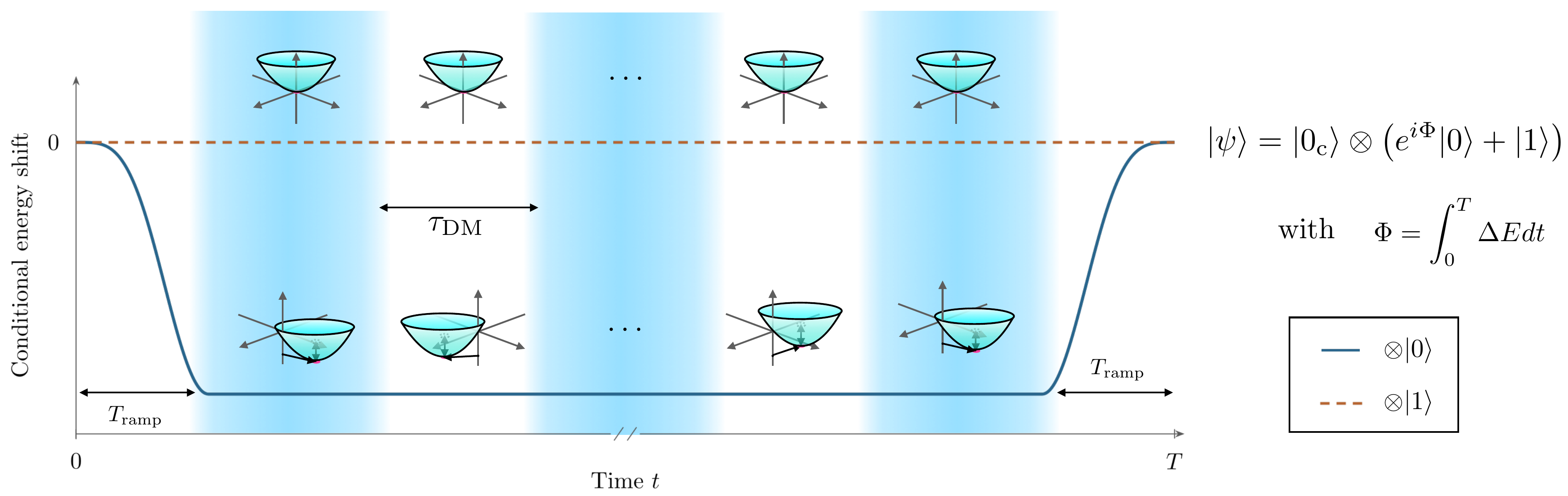}
    \caption{A sketch of our protocol for accumulating a dark matter-induced phase over multiple dark matter coherence times (represented by blue shades). For simplicity, this example uses two branches, which are controlled by a register qubit. The system is initialized in the state $|\psi_i\rangle=|0_c\rangle\otimes\left(|0\rangle+ |1\rangle \right)/\sqrt{2}$. The dark matter signal is ramped on, say by shifting to small detuning, or as described in Section~\ref{sub:Ramping}, conditioned on the qubit being in $\ket{0}$, and is then left for a time $T_\mathrm{free} \gg \tau$. During this time the cavity state evolves adiabatically, changing its displacement direction on a scale of $\tau_\mathrm{DM}$. Nonetheless, the energy shift has a non-zero time average over the full exposure, allowing the cavity to coherently accumulate a large repeatable phase. The signal is then ramped off adiabatically, allowing the phase to be encoded in the ancilla qubit alone. }
    \label{fig:clock-branches}
\end{figure*}

\subsection{Effective Signal Ramping}\label{sub:Ramping}

Although the adiabaticity condition above ensures that a cavity which is at the instantaneous Hamiltonian extremum at one time will remain at that extremum as the dark matter state evolves, it does not answer the question of how one first gets into the instantaneous extremum. In principle, one way to ensure this would be to start the cavity in vacuum, and simultaneously to make sure that the instantaneous extremum of its potential also begins there, before slowly evolving that minimum to $-\sigE(t)/\delta_\bk$; one could thus transition slowly from the left to the right panel of Figure~\ref{fig:shift}.

This would be straightforward for a signal under our control, but the presence of dark matter in the cavity cannot generally be turned on and off. We can, however, continuously change the cavity's sensitivity to the dark matter, which has the same effect. There are various ways one could imagine doing this: one could, for example, dramatically shift the cavity frequency to ensure a very large detuning that renders the phase $\Phi \propto \sigErms^2/\delta_\bk$ negligible for all bins, and then slowly reduce this detuning contribution to zero to recover the behavior above (although this can create a problem discussed in Section~\ref{sub:ImperfectDecoupling}). The specific strategy that we will present (following Ref.~\cite{Allen:2025hdx}) is dynamical decoupling, which is conceptually somewhat cleaner, but we note that it may or may not be the optimal strategy subject to the hardware constraints of a particular detector.

To implement dynamical decoupling, we apply repeated $\pi$
rotations of the cavity,
\begin{align}
    \Pi=e^{i\pi\hat a^\dagger\hat a},
    \qquad
    \Pi^\dagger\hat a\Pi=-\hat a.
\end{align}
Each rotation reverses the sign of the dark matter-induced drive while leaving the detuning term unchanged. In a frame following these rotations, the drive therefore alternates in sign. Equal intervals with opposite signs cancel its effect to leading order in the pulse-cycle duration; we consider sub-leading corrections in Section~\ref{sub:ImperfectDecoupling}.

A control qubit with dispersive coupling $\chi_\DD$ can implement each rotation by activating its cavity-frequency shift for a duration $T_\DD=\pi/\chi_\DD$~\cite{Sun2014}. We choose the pulse-cycle duration $\Delta t$ to satisfy $T_\DD\ll\Delta t\ll\min(1/\dchar,\tau,T_\ramp)$, where $T_\ramp$ is the duration of the effective signal ramp~\cite{VitaliTombesi1999}. This hierarchy is possible when
\begin{align}
    \chi_\DD \gg \dchar, \, \frac{1}{\tau}, \, \frac{1}{T_\ramp}.
\end{align}
Here $\dchar$ is the characteristic detuning of the signal bins, defined in Section~\ref{sub:Binning}.

To obtain a nonzero effective drive, we make alternating intervals between pulses unequal. Slowly varying this imbalance controls the cycle-averaged drive amplitude. The effective Hamiltonian in Eq.~\eqref{eq:effHcav} then becomes
\begin{align}
    H_\DD(t) = \delta_\bk \hat a^\dagger\hat a + \eta(t)\left( \sigE(t) \hat{a}^\dagger + \sigE(t)^* \hat{a} \right), \label{eq:HcavDD}
\end{align}
where $\eta(t)\in[0,1]$ is the effective coupling envelope. We begin with $\eta=0$, smoothly increase it to $\eta=1$ over a time $T_\ramp\ll T_\free$, hold it there (i.e. do not apply $\Pi$ pulses) for $T_\free$, and then ramp back to $\eta=0$ over another $T_\ramp$.

For detunings and ramp times that satisfy the adiabatic conditions, the cavity follows the displaced state connected to vacuum. In the adiabatic limit, it returns to the vacuum for every $\bk$, while the register components acquire the phases $\Phi_\bk$. The final register and cavity are therefore no longer entangled after the ramp-off~\cite{CrossGambetta2015,Paik2016}~(see also Eq.~\eqref{eq:NoisyOracleEvolution} below), leaving the accumulated phases available for subsequent quantum processing.

We illustrate a simple example of this evolution in Figure~\ref{fig:clock-branches}, showing the case in which there is only one register qubit, which determines if the effective dark matter interaction is ``on'' (small detuning) or ``off'' (large detuning). Because the ramp-off process brings the system adiabatically to the dark matter-decoupled state, the phase can be read out as a $Z$ rotation on the qubit; this is reminiscent of an atomic clock, which reads out a phase of energy-difference times time using interference. The multi-qubit system described above can implement a similar protocol over $2^{n_q}$ branches, rather than just $2$. Only branches with a small (but generally non-zero) detuning will accumulate a large phase. In the next Section~\ref{sec:GroverSearch}, we will show how to adapt this protocol into an oracle that can be used in a Grover search.

\section{A Grover Search for Dark Matter}\label{sec:GroverSearch}

\begin{figure*}[t]
  \centering
  \includegraphics[width=0.9\textwidth]{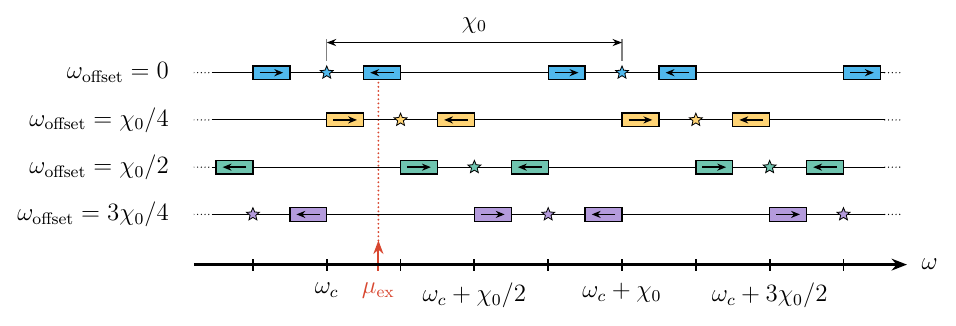}
  \caption{An illustration of our binning scheme. Each Grover search leaves deliberate gaps between frequency bins to localize the oracle phase to one bin. Bins with $\omof = 0$, $\chi_0/4$, $\chi_0/2$, and $3\chi_0/4$ are shown in blue, yellow, green, and purple, respectively, and the choice between them is made once for a full Grover search using an offset qubit. Cavity mode frequencies corresponding to different register states $\ket{\bk}$ are shown as stars, and frequency bins are shown as rectangles with arrows pointing towards the corresponding cavity frequency. For illustration, we add an example signal frequency $\mu_\mathrm{ex}$ marked by a red arrow. For this example the signal bin $\bk^*$ is the blue ($\omof = 0$) $\omega_\bk$---not the nearest central frequency, which has $\omof = \chi_0/4$.}
  \label{fig:BinningSketch}
\end{figure*}

Over the last two sections we have described a way to engineer a system that evolves over a time $\sim T_\free$ as
\begin{align}
    \left( \sum_{\bk = 0}^{2^{n_q}-1} \ket{\bk} \right) \otimes \ket{0_c} \longrightarrow \left( \sum_{\bk = 0}^{2^{n_q}-1} e^{i\Phi_\bk} \ket{\bk} \right) \otimes \ket{0_c}, \label{eq:NoisyOracleEvolution}
\end{align}
where $\ket{0_c}$ is the cavity vacuum and we have omitted the normalization for compactness. What we will ultimately require, in order to have a standard Grover oracle for our eventual use of Grover's algorithm, is replacing the phases $\Phi_\bk$ with $\pi \delta_{\bk,\bk^*}$, where $\bk^*$ is the bin containing the signal. A replacement of this type will be the goal of the first three subsections below: In the first, we will show a possible binning strategy that will restrict the signal phase primarily to one bin and to a limited range of possible values. In the second and third subsections, we will discuss digitizing this phase, ultimately mapping it to either 0 or $\pi$ depending on whether it is below or above some threshold.  We emphasize, however, that the binning we present in Section~\ref{sub:Binning} below is merely an example to illustrate that the (different) binning strategy in Ref.~\cite{Allen:2025hdx} is not the only option; we leave finding optimal binning strategies, which requires a precise treatment of order-one numbers, to future work.

\subsection{Binning}\label{sub:Binning}

We begin by setting the ranges of possible $\delta_\bk$ values, both in our signal bin and in non-signal  bins. We accomplish this using a somewhat non-trivial binning strategy, in which our signal will be looked for not in the bin where it has the smallest detuning $\delta_\bk$, but in the one where its detuning is second-smallest. Note that our binning strategy is different from the one used in Ref.~\cite{Allen:2025hdx}, allowing us to more quickly search for signals above the minimum detectable signal strength (see Section~\ref{sub:Rescan}). This does not parametrically change the detector's sensitivity scaling, but can reduce total observation time by a constant factor of $\sim 10$, as we discuss in Section~\ref{sub:Rescan}. An optimal binning strategy is an interesting direction for exploration in future work; we discuss some aspects of this optimization at the end of Section~\ref{sub:Rescan}.

We illustrate our binning in Fig.~\ref{fig:BinningSketch}. It relies on four separate, full Grover searches (i.e. everything we describe up to and including Section~\ref{sub:GroversAlg}, as well as the signal confirmation operations of Section~\ref{sub:Rescan}), corresponding to each combination of the control qubit states, i.e. to $\omof = 0$, $\chi_0/4$, $\chi_0/2$, and $3\chi_0/4$. The four runs are thus identical up to this global frequency shift, which appears in quarters of our usual adjacent-$\bk$ spacing. (In fact we will later discuss using more Grover searches than this, with other parameters changed, but for now it is sufficient to consider four of them.)

Within each run, we have $\Nbins = 2^{n_q}$ bins, corresponding to the $\Nbins$ different basis states of the qubits, or, equivalently, the $\Nbins$ different state components in Eq.~\eqref{eq:NoisyOracleEvolution}. Each of these bins is characterized by a central cavity frequency $\omega_\bk$, and will correspond, at the smallest coupling we try to detect, to looking for dark matter with mass in one of two ranges:
\begin{align}
\begin{split}
    \mu &\in \left[ \omega_\bk - \frac{1}{4}\chi_0 , \omega_\bk - \frac{1}{8}\chi_0 \right] \qquad {\rm or} \\
    \mu &\in \left[ \omega_\bk + \frac{1}{8}\chi_0 , \omega_\bk + \frac{1}{4}\chi_0 \right].
\end{split}
\end{align}
Equivalently, each bin will correspond to searching for detunings
\begin{align}
    \frac{\chi_0}{8} \leq |\delta_\bk| \leq \frac{\chi_0}{4}.
\end{align}
The center of this range is the characteristic detuning we have referred to until now, $\dchar \sim \chi_0/6$. Every other bin at the same $\omof$ setting will have a detuning of at least $3\chi_0/4$.

Note a very peculiar feature of this binning scheme: we look for dark matter not in the bin with the smallest detuning (between all four runs), but in the bin with the second-smallest (although it will be the smallest detuning among the bins at its $\omof$ value). This bin will be called~$\bk^*$. We will return to what happens in the genuine smallest-detuning bin below.

We noted in Section~\ref{sub:FreeEvolution} that we could safely assume that the cavity would adiabatically follow its instantaneous co-rotating extremal state so long as $\delta_\bk \tau \gg 1$. Since the smallest detuning we consider here is $\chi_0/8$, we therefore choose our bin spacing to be
\begin{align}
    \chi_0 = \frac{c}{\tau},
\end{align}
where $c$ is some constant, likely of order several to ensure adiabaticity, but which should ultimately be found by numerically simulating the system evolution and then optimizing. We leave doing so to future work, since this will no doubt depend on the exact design of the experiment, whereas we are presenting only a toy model. Note that, as we discussed below Eq.~\eqref{eq:AdiabaticityDetuningCondition}, our protocol sets a stronger condition on the minimum well-behaved signal detuning than the perfect-coherence approach in Ref.~\cite{Allen:2025hdx}; correspondingly our bins are widened from $\sim \sigErms$ to $\sim \tau^{-1}$.

During that run the phase acquired by $\ket{\bk^*}$ is
\begin{align}
    |\Phi_{\bk^*}| \in \left[ \frac{4\sigErms^2 T_\free \tau}{c} , \frac{8\sigErms^2 T_\free \tau}{c} \right].
\end{align}
Setting
\begin{align}
    T_\free = \frac{c\pi}{20\sigEmin^2\tau} \label{eq:TfreeVal}
\end{align}
where $\sigEmin$ is the smallest value of $\sigErms$ that we will aim to reliably detect, and defining
\begin{align}
    r = \frac{\sigErms^2}{\sigEmin^2} \geq 1  
\end{align}
as a possible enhancement of the actual $\sigErms$ with respect to the target $\sigEmin$, we then have
\begin{align}
    |\Phi_{\bk^*}| \in \left[ \frac{\pi}{5} r, \frac{2\pi}{5} r \right]. \label{eq:PhiStarRange}
\end{align}
All of the other bins during that run will have
\begin{align}
    |\Phi_{\bk \neq \bk^*}| \leq \frac{\pi}{15} r. \label{eq:PhiFarRange}
\end{align}
This separation in the magnitude of $\Phi$ between signal ($\bk^*$) and non-signal (all other $\bk)$ bins will allow us to digitize the phase---converting it to a phase $\pi$ for the signal bin and $0$ otherwise---as we show in Sections~\ref{sub:QSPGeneral} and~\ref{sub:QSPforDM}.

There are two remaining caveats. First, what happens during the other runs (i.e. other $\omof$ values)? The answer is simply that it doesn't matter: it is entirely possible that the run which contains the least-detuned bin of all sees a signal in that bin and we find it there; it is also possible that, because the detuning is too small, the phase ends up being so large that we do not see it (e.g. $\Phi_\bk = 2\pi$, which of course cannot be detected), or that the small detuning results in a non-zero occupation of the cavity and a subsequent decay that decoheres the detector state, or various other possibilities. Similarly it may be that we see the signal at some other $\omof$ value for $r>1$, but this is not guaranteed. None of these matter for us, however; we rely only on seeing the signal during the ``safe,'' more-detuned run.

The second caveat is how we handle the unknown signal strength, i.e. the unknown value of $\sigErms$ or, equivalently, $r$.
Treating $r>1$ is more involved, and there are various ways to address it; we present one option in some detail, and mention others, in Section~\ref{sub:Rescan}.

\subsection{Quantum Signal Processing}\label{sub:QSPGeneral}

The phase accumulator in Equation~\eqref{eq:NoisyOracleEvolution} is a ``noisy oracle'': it produces an order-one phase in one bin (or two) for couplings $\gtrsim\sigEmin$. In order to turn this into a ``clean'' Grover oracle, with a phase $\pi$ for $\bk = \bk^*$ and $0$ otherwise, we employ quantum signal processing (QSP), as in Ref.~\cite{Allen:2025hdx}. In this section, we will explain what QSP is and how it can be applied to our setup, but we treat it essentially as a black box; see instead Ref.~\cite{Haah:2019pes} for the Laurent-polynomial form applicable to our design, or Ref.~\cite{Martyn:2021eaf} for a more pedagogical introduction. For illustrative purposes, we also present an example of a specific QSP sequence in Appendix~\ref{app:QSP}.

Suppose that one has a single qubit and the ability to repeatedly apply an operation
\begin{align}
    U_\Phi = 
    \begin{pmatrix}
        e^{i\Phi} & 0 \\
        0 & e^{-i\Phi}
    \end{pmatrix} \label{eq:QSPSignalForm}
\end{align}
to it, where $\Phi$ is some unknown angle (which will eventually be our $\Phi_\bk$\footnote{This is why, in Section~\ref{sub:Binning}, we set acceptance windows around $\Phi = \pi/2$ rather than $\pi$, so that the ancilla rotation has relative phase $\pi$. QSP subsequently converts all accepted phases into a register phase of $\pi$.}), as well as to apply arbitrary, controlled rotations $R$ to the qubit. For us, applying $U_\Phi$ will amount to exposing the cavity to dark matter, although only after we make some changes to the protocol we described in Section~\ref{sec:CoherentPhaseAcc}. In QSP, one chooses to alternate applications of $U_\Phi$ and some chosen $\Phi$-independent rotations such that, after $N_{\rm QSP}$ iterations---i.e. after applying
\begin{align}
    U_{\rm QSP} = R_{N_{\rm QSP}} U_\Phi R_{N_{\rm QSP}-1} U_\Phi \, ... \, R_1 U_\Phi R_0
\end{align}
to a chosen initial state (say $\ket{0}$)---one has achieved the unitary evolution
\begin{subequations}
\begin{align}
    \ket{0} &\longrightarrow P(e^{i\Phi})\ket{0} + Q(e^{i\Phi})\ket{1} \\
    \ket{1} &\longrightarrow -Q(e^{i\Phi})^*\ket{0} + P(e^{i\Phi})^*\ket{1}.
\end{align}
\end{subequations}
Here $P$ and $Q$ are finite Fourier sums, or equivalently Laurent polynomials in $e^{i\Phi}$ of degree $\leq N_\QSP$, i.e.
\begin{subequations}
\begin{align}
    P\left(e^{i\Phi'}\right) &= \sum_{\substack{m=-N_\QSP\\m\equiv N_\QSP\ ({\rm mod}\,2)}}^{N_\QSP} p_m e^{im\Phi'} \\ 
    Q\left(e^{i\Phi'}\right) &= \sum_{\substack{m=-N_\QSP\\m\equiv N_\QSP\ ({\rm mod}\,2)}}^{N_\QSP} q_m e^{im\Phi'}.
\end{align}
\end{subequations}
whose coefficients are constrained by unitarity:
\begin{align}
    \left|P\left(e^{i\Phi'}\right)\right|^2 + \left|Q\left(e^{i\Phi'}\right)\right|^2 = 1
\end{align}
for all phases $\Phi'$. We will choose $N_\QSP$ to be even, giving $\pi$-periodic functions.

The evolution that we would like is given by
\begin{align}
\begin{split}
    P_{\rm ideal}\left(e^{i\Phi}\right) &= 
    \begin{cases}
        -1 \qquad \frac{\pi}{5} \leq \bar{\Phi} \leq \frac{4\pi}{5} \\
        1 \hphantom{-}\qquad {\rm otherwise}
    \end{cases} \\
    Q_{\rm ideal}\left(e^{i\Phi}\right) &= 0,
\end{split}   
\end{align}
where we define $\bar{\Phi}$ to be $\Phi \pmod{\pi}$, so $\bar{\Phi} \in [0,\pi)$. The lower limit $\pi/5$ is simply the lower limit of the interval in Eq.~\eqref{eq:PhiStarRange}, but the upper limit has been increased from $2\pi/5$ to $4\pi/5$ to handle $r>1$, as we discuss in Section~\ref{sub:Rescan}. The step function $P_{\rm ideal}$ is of course not such a Laurent polynomial, but it can be approximated by one, with degree~\cite{Gilyen:2018khw, Low:2016jxt, Haah:2019pes}
\begin{align}
    N_{\rm QSP} = O\left( \frac{1}{\Delta} \ln\pfrac{1}{\xi} \right)
\end{align}
where $\Delta$ is the width of the transition between the $\pm1$ phases and $\xi$ is a maximum error (outside of the transition regions),
\begin{align}
    \left| P_{\rm QSP} - P_{\rm ideal} \right| \leq \xi
\end{align}
and thus
\begin{align}
    \left| Q_{\rm QSP} \right|^2 \leq 2\xi
\end{align}
by unitarity. For concreteness, in this work we will assume a QSP that gives
\begin{align}
\begin{split}
    P_{\rm QSP}\left(e^{i\Phi}\right) &\approx 
    \begin{cases}
        -1 \qquad \frac{\pi}{5} \leq \bar{\Phi} \leq \frac{4\pi}{5} \\
        1 \hphantom{-}\qquad \bar{\Phi} \leq \frac{2\pi}{15} {\rm~or~} \frac{13\pi}{15} \leq \bar{\Phi}
    \end{cases}
\end{split} 
\end{align}
(and thus $Q_\QSP(e^{i\Phi}) \approx 0$ in both listed regimes), with some continuous transition in the $2\pi/15 < \bar{\Phi} < \pi/5$ and $4\pi/5 < \bar{\Phi} < 13\pi/15$ regimes. The amplitude bound above implies a unitary error at most $\sqrt{2\xi}$ per digitized phase; since we will eventually perform $\sim\sqrt{\Nbins}$ Grover iterations, this ensures a small total error for $\xi \ll 1/\Nbins$. We emphasize, however, that this is merely an example, and we leave the task of numerical optimization to find a specific QSP appropriate to an experiment's error tolerance to future work.

Unfortunately, generating such a QSP requires access to phases of both signs: our protocol in Section~\ref{sec:CoherentPhaseAcc} allows us to apply phases of $+\Phi_\bk$, but we will additionally need access to $-\Phi_\bk$ (or at least we are not aware of any way to generate an appropriate QSP without access to the opposite sign). We present one example approach to generating $-\Phi_\bk$ in the next subsection.

\subsection{Phase Digitization for Dark Matter}\label{sub:QSPforDM}

To achieve both possible phase signs, we will effectively turn our cavity into a qubit, at which point our strategy is closely analogous to that of Ref.~\cite{Allen:2025hdx}. In principle, this can be done using a photon blockade that splits the $\ket{2_c}$ Fock level into two levels that are both far-detuned from dark matter-induced transitions based on a photon-number-selective dressing by a driven auxiliary transmon~\cite{Chakram:2020eky}; it can also be done by introducing a Kerr non-linearity $H_{\rm Kerr} =  \kappa_{\rm Kerr} \hat{a}^\dagger \hat{a} (\hat{a}^\dagger \hat{a}-1)$ to the cavity Hamiltonian using a Josephson-junction-induced cavity anharmonicity~\cite{Kirchmair:2012rsm}, shifting all levels with 2 or more photons far out of resonance. In both cases, the important effect is that one can, with sufficiently strong terms, eliminate the effect of these higher levels from the behavior of the cavity's $\ket{0_c}$ and $\ket{1_c}$ Fock levels. Making these terms sufficiently large is impractical with present technology, as we discuss in Section~\ref{sec:TechReqs}. We suggest these simply as a proof-of-principle for our approach, to show that difficulty of applying Grover's algorithm to dark matter search is technological, rather than a fundamental limitation of dark matter.

We can understand the effect of eliminating the higher Fock levels on our system by treating the $\ket{0_c}$ and $\ket{1_c}$ levels separately. The evolution of a cavity state starting in $\ket{0_c}$, i.e. in the vacuum, as in Section~\ref{sec:CoherentPhaseAcc}, is essentially unchanged: the effect of dark matter, Eq.~\eqref{eq:HDM} is linear in $\hat{a}$ and $\hat{a}^\dagger$, and thus it couples only adjacent Fock levels at lowest order in perturbation theory; since the effect of dark matter is generally tiny, the higher-order contributions should be negligible for a typical dark matter search (although of course one should verify this for any particular experiment). Thus that vacuum state, subject to the coherent phase accumulation of Section~\ref{sec:CoherentPhaseAcc}, will acquire, at leading order, the phase
\begin{align}
    \Phi_\bk = \frac{\sigErms^2 T_\free}{\delta_\bk}
\end{align}
(up to small ramp corrections) calculated previously. Note that the trajectory of the cavity state is no longer exactly a coherent state, but the accumulated phase agrees with the harmonic result at leading order.

The behavior of states starting in $\ket{1_c}$---which we will need below---is more meaningfully modified by the removal of the higher Fock levels, however, since that state is coupled to $\ket{2_c}$ by dark matter even at leading order. A simple observation allows us to avoid working this behavior out in detail, however: since $H_{\rm DM}$ is fully off-diagonal in Fock number, it does not change the trace of the Hamiltonian describing the $\{\ket{0_c}, \, \ket{1_c}\}$ subspace of the cavity (in the ideal two-level limit), and thus it cannot change the sum of those two states energies. A state beginning in $\ket{1_c}$ is thus guaranteed to acquire the same signal-induced phase as one starting in $\ket{0_c}$, but with the opposite sign, subject to our usual adiabaticity assumptions. This will allow us to generate both signs of $\Phi_\bk$ below.

To implement this in our setup, we introduce one additional ancilla qubit, whose coupling to the cavity is activated only during the operations described below, and initiate it in the $\ket{+_A} = (\ket{0_A}+\ket{1_A})/\sqrt{2}$ state. We now perform two operations based on this qubit: prior to ramping the dark matter signal on, we apply what is essentially a CX gate to the cavity, conditional on the ancilla state:
\begin{align}
\begin{split}
    {\rm CX} = & \ket{0_A}\bra{0_A} \otimes I_c \\
    \hphantom{=} & + \ket{1_A}\bra{1_A} \otimes \left( \ket{0_c}\bra{1_c} + \ket{1_c} \bra{0_c} \right)
\end{split}
\end{align}
with $I_c$ the identity operator on the cavity, up to higher-Fock state behavior that does not matter here. This could be implemented using, for example, a tunable dispersive coupling to a control qubit~\cite{Allman:2010xlc, Chen:2014xki} combined with rotations of the cavity qubit, which can be produced by ordinary coherent cavity drives once the higher Fock levels are eliminated. The result is that the cavity coupled to one ancilla state will begin in the vacuum, but the other will begin with one photon in it, and thus the two will acquire opposite phases over the course of the evolution described in Section~\ref{sec:CoherentPhaseAcc}. We then apply the same CX gate after the signal is ramped off, returning the cavity to vacuum for both ancilla states, but retaining the phase difference on the ancilla. The result is the evolution
\begin{align}
\begin{split}
    \sum_{\bk = 0}^{\Nbins-1} \ket{\bk} \ket{0_c} \ket{0_A} &\longrightarrow \sum_{\bk = 0}^{\Nbins-1} e^{i\Phi_\bk} \ket{\bk} \ket{0_c} \ket{0_A} \\
    \sum_{\bk = 0}^{\Nbins-1} \ket{\bk} \ket{0_c} \ket{1_A} &\longrightarrow \sum_{\bk = 0}^{\Nbins-1} e^{-i(\nu_\bk + \Phi_\bk)} \ket{\bk} \ket{0_c} \ket{1_A}
\end{split}
\end{align}
where $\nu_\bk$ is the phase acquired by the $\ket{1_A}$-coupled state even in the absence of dark matter due to its higher energy. This energy is a function solely of known cavity properties, however, and so it can be corrected away using a $\bk$-dependent ancilla rotation
\begin{subequations}
\begin{align}
    \ket{\bk} \ket{0_A} &\longrightarrow \ket{\bk} \ket{0_A} \\
    \ket{\bk} \ket{1_A} &\longrightarrow \ket{\bk} e^{+i\nu_\bk} \ket{1_A}.
\end{align}
\end{subequations}
This finally gives us
\begin{align}
\begin{split}
    \sum_{\bk = 0}^{\Nbins-1} \ket{\bk} \ket{0_c} \ket{0_A} &\longrightarrow \sum_{\bk = 0}^{\Nbins-1} e^{i\Phi_\bk} \ket{\bk} \ket{0_c} \ket{0_A} \\
    \sum_{\bk = 0}^{\Nbins-1} \ket{\bk} \ket{0_c} \ket{1_A} &\longrightarrow \sum_{\bk = 0}^{\Nbins-1} e^{-i\Phi_\bk} \ket{\bk} \ket{0_c} \ket{1_A}.
\end{split}
\end{align}

We can then implement our needed QSP simply by alternating intervals of free evolution, including the ramping on and off and the operations above, with appropriate rotations of the ancilla. Together these give us, after $N_\QSP$ repetitions,
\begin{align}
    \sum_{\bk = 0}^{\Nbins-1} \ket{\bk} \ket{0_c} \ket{+_A} \longrightarrow \sum_{\bk = 0}^{\Nbins-1} e^{i\pi\delta_{\bk,\bk^*}} \ket{\bk} \ket{0_c} \ket{+_A}, \label{eq:CleanOracle}
\end{align}
which is precisely what we will need for Grover's algorithm.

\subsection{Grover's Algorithm}\label{sub:GroversAlg}

\begin{figure*}
    \centering
    \includegraphics[width=1.0\linewidth]{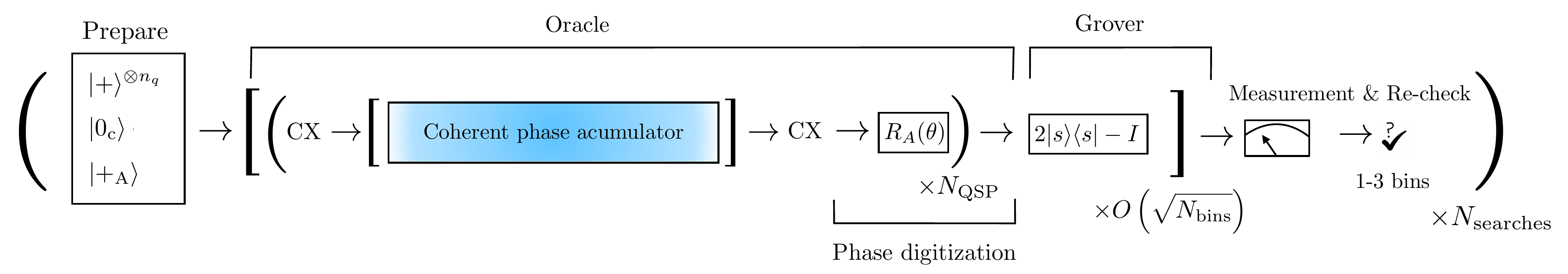}
    \caption{A sketch of our protocol: one prepares a superposition state, accumulates dark matter phase over many dark matter coherence times, digitizes the phase (via $N_\mathrm{QSP}$ repetitions), Grover searches via amplitude amplification ($\sqrt{\Nbins}$ repetitions), and finally measures and re-checks candidate signals. The procedure is repeated to cover frequency bins with four values of $\omof$ separately, as well as to search for different ranges of signal strengths $\sigErms$.}
   \label{fig:OperationSequence}
\end{figure*}

At this point we have constructed a ``clean'' Grover oracle \eqref{eq:CleanOracle} and thus we can apply Grover's algorithm to identify the bin $\bk^*$. We briefly summarize Grover's  algorithm here, but see Ref.~\cite{GroverOriginal} for the original algorithm and e.g. Ref.~\cite{Boyer:1996zf} for its extension to multiple bins with phases of $\pi$ (hereafter ``marked'' bins), which we will briefly discuss later. We begin with the single-marked-bin case.

At the beginning of the Grover search (i.e. before our first QSP cycle, or even our first signal ramp on), we have prepared the state
\begin{align}
    \ket{\psi_0} = \frac{1}{\sqrt{\Nbins}} \sum_\bk \ket{\bk} \ket{0_c} \ket{+_A}.
\end{align}
For the rest of this section we will omit the cavity and ancilla states, since they will not be involved in the dynamics. Then write
\begin{align}
\begin{split}
    \ket{\psi_0} &= \frac{1}{\sqrt{\Nbins}} \ket{\bk^*} + \sqrt{\frac{\Nbins-1}{\Nbins}} \ket{\perp} \\
    &= \sin\theta_0 \ket{\bk^*} + \cos\theta_0 \ket{\perp}
\end{split}
\end{align}
where
\begin{align}
    \ket{\perp} = \frac{1}{\sqrt{\Nbins-1}} \sum_{\bk \neq \bk^*} \ket{\bk}.
\end{align}
and $\theta_0 = \sin^{-1}(1/\sqrt{\Nbins}) \approx 1/\sqrt{\Nbins}$.

We will now repeatedly apply the operator
\begin{align}
    G = DO
\end{align}
to our state, where
\begin{align}
    D = 2\ket{\psi_0}\bra{\psi_0} - 1
\end{align}
and $O$ is the QSP-cleaned oracle of Eq.~\eqref{eq:CleanOracle}. After one iteration, this gives
\begin{align}
\begin{split}
    G\ket{\psi_0} &= \sin\left(3\theta_0\right) \ket{\bk^*} + \cos\left(3\theta_0\right) \ket{\perp},
\end{split}
\end{align}
i.e. a rotation by angle $2\theta_0$ towards the state $\ket{\bk^*}$. This is easily shown to generalize: each successive application of the same operator $G$ rotates the state by another $2\theta_0$. This process is known as ``amplitude amplification''~\cite{Brassard2002}. Repeating this process $N_{\rm Grover} \approx \pi\sqrt{\Nbins}/4$ times gives
\begin{align}
    G^{N_{\rm Grover}} \ket{\psi_0} \approx \ket{\bk^*},
\end{align}
and thus a final measurement of each register qubit tells us the signal bin, $\bk^*$.

In the presence of two marked bins (i.e. two bins that receive a $-1$ phase from the QSP; this will appear briefly in Section~\ref{sec:GroverSearch}), Grover's algorithm can be used essentially unchanged, except that it now takes only $N'_{\rm Grover} \approx \pi\sqrt{\Nbins/2}/4$ iterations, after which a measurement will give either of the two bins, with essentially equal probability~\cite{Boyer:1996zf}. Note, however, that these differing numbers of Grover iterations mean that one must perform the two searches separately.

\subsection{The Full Protocol, Signal Confirmation, and Repeated Scans}\label{sub:Rescan}

We can now finally assemble all of the components of our Grover search, including some necessary follow-up measurements. For simplicity, let us first restrict to $r=1$, i.e. to a search for a particular value of $\sigErms = \sigEmin$, before we explain how one handles $r>1$. A single Grover search, which gives one candidate signal bin, is illustrated in Fig.~\ref{fig:OperationSequence}: 
\begin{enumerate}
    \item Prepare the cavity in the vacuum $\ket{0_c}$ while operating the dynamical decoupling to keep its extremal state at the origin; set the register qubits in $\ket{+_k}$ for all $k$; and set the ancilla to $\ket{+_A}$. Also set the appropriate $\omof$ value for this search; say $\omof = 0$ first.
    
    \item Next, apply CX to the cavity conditional on the ancilla state, and then gradually turn off the dynamical decoupling over a time $T_\ramp$, slowly shifting the cavity Fock states to their dark matter-dressed states.
    
    \item Allow the cavity to freely evolve for a time $T_\free = c\pi/(20\sigEmin^2\tau)$.
    
    \item Finally, gradually turn the dynamical decoupling back on, adiabatically bringing the cavity branches back to $\ket{0_c}$ and $\ket{1_c}$, and again apply CX to the cavity conditional on the ancilla state, as well as the signal-independent phase correction. The result is the phase acquisition
    \par\noindent
    \begin{minipage}{\dimexpr\columnwidth-\leftmargin-\rightmargin\relax}
    \setlength{\abovedisplayskip}{0pt}
    \setlength{\abovedisplayshortskip}{0pt}
    \begin{subequations}
    \begin{align}
        \ket{\bk}\ket{0_c}\ket{0_A} &\longrightarrow e^{i\Phi_\bk} \ket{\bk}\ket{0_c}\ket{0_A} \\
        \ket{\bk}\ket{0_c}\ket{1_A} &\longrightarrow e^{-i\Phi_\bk} \ket{\bk}\ket{0_c}\ket{1_A},
    \end{align}
    \end{subequations}
    \end{minipage}
    \par\noindent
    where
    \par\noindent
    \begin{minipage}{\dimexpr\columnwidth-\leftmargin-\rightmargin\relax}
    \setlength{\abovedisplayskip}{0pt}
    \setlength{\abovedisplayshortskip}{0pt}
    \begin{align}
        \Phi_\bk = \frac{\pi}{20} \frac{c}{\delta_\bk\tau},
    \end{align}
    \end{minipage}
    \par\noindent
    which is order-one for $\delta_\bk \sim \delta^*$. From a quantum computational standpoint, steps $2-4$ can be viewed as acting on the register and ancilla with a repeatable $\bk$-dependent phase gate.
    
    \item Now apply the chosen QSP rotation to the ancilla, and repeat the above steps (starting from the CX gate) with additional QSP rotations for a total of $N_{\rm QSP}$ iterations. This gives
    \par\noindent
    \begin{minipage}{\dimexpr\columnwidth-\leftmargin-\rightmargin\relax}
    \setlength{\abovedisplayskip}{0pt}
    \setlength{\abovedisplayshortskip}{0pt}
    \begin{align}
        \ket{\bk}\ket{0_c}\ket{+_A} \longrightarrow e^{i\pi\delta_{\bk,\bk^*}} \ket{\bk}\ket{0_c}\ket{+_A},
    \end{align}
    \end{minipage}
    \par\noindent
    up to an exponentially-suppressed gate synthesis error.
    
    \item The above (starting from the initial QSP rotation and the CX gate) forms one Grover oracle; now apply the Grover operator
    \par\noindent
    \begin{minipage}{\dimexpr\columnwidth-\leftmargin-\rightmargin\relax}
    \setlength{\abovedisplayskip}{0pt}
    \setlength{\abovedisplayshortskip}{0pt}
    \begin{align}
        D = 2\ket{\psi_0}\bra{\psi_0} - 1,
    \end{align}
    \end{minipage}
    \par\noindent
    giving one Grover iteration. This should be repeated $\pi\sqrt{\Nbins}/4$ times.
    
    \item At the end of those Grover iterations, measure the state of the $n_q$ register qubits. This gives a single candidate bin $\bk$.
\end{enumerate}

Note, however, that---even in the absence of any errors---this bin is not guaranteed to be dark matter: the Grover search will \emph{always} give some bin as output, even if there was no signal at all. One should therefore proceed as follows: 
\begin{enumerate}
    \setcounter{enumi}{7}
    \item Perform a non-Grover search of that one candidate bin. Although there are various ways to do so, one is to simply employ the strategy in Section~\ref{sec:CoherentPhaseAcc}, with $n_q = 1$ and the cavity set to be at the candidate frequency for one register qubit state and far detuned for the other. No Grover search is needed, and the QSP is likely inferior to simply measuring a few times in order to handle phases that are large but not exactly $\pi$.
   
    \item Repeat the Grover search with the other three $\omof$ values, checking the other three quarters of the frequency range.
\end{enumerate}

Together, these four Grover searches plus their subsequent individual-bin checks form a complete search of one $\sigErms = \sigEmin$ value, i.e. the search at $r=1$. A signal with $r=1$ should give a candidate signal at least during the Grover run when the signal is in a bin's assigned frequency range, at which point it can be confirmed; whether it appears in any of the other runs does not matter.

We now consider the result of this procedure at higher signal coupling,~$r>1$. Recall that our phases for the true signal bin $\bk^*$ and all other bins, are given by (see Eqs.~\eqref{eq:PhiStarRange}-\eqref{eq:PhiFarRange})
\begin{subequations}
\begin{align}
    |\Phi_{\bk^*}| &\in \left[ \frac{\pi}{5} r, \frac{2\pi}{5} r \right] \\
    |\Phi_{\bk \neq \bk^*}| &\leq \frac{\pi}{15} r.
\end{align}
\end{subequations}
Thus, for $1 < r < 2$, it is both the case that
\begin{itemize}
    \item The phase in the $\bk^*$ bin is between $\pi/5$ and $4\pi/5$, i.e. in the $P(e^{i\Phi}) \approx -1$ range of the QSP phase digitizer; and
    \item The phase in all bins except $\bk^*$ is less than $2\pi/15$, i.e. in the $P(e^{i\Phi}) \approx 1$ range of the QSP phase digitizer.
\end{itemize}
The search above has therefore already acted as a search for all $r$ values from $1$ to $2$. At larger $r$, however, even bins other than $\bk^*$ could begin to acquire non-negligible phases, and thus the search above may not succeed.

In order to search for signals at larger $r$, then (e.g. to do a search for all signal above some minimum threshold), one simply repeats the experiment with a higher target coupling: thus if one has previously excluded all $\sigErms$ above some $\sigEmax$ (with other experiments), one can perform a search with $\sigEmin$ set to $\sigEmax/\sqrt{2}$, and then one at $\sigEmax/2$, and so on until one has reached one's desired sensitivity. Crucially, although this is a logarithmic (in range of $\sigErms$ values covered) number of repetitions, the time cost of doing this is only a factor of 2 (compared to the smallest-$\sigEmin$ search), since a search conducted at twice the $\sigEmin^2$ value takes only half as long (see Eq.~\eqref{eq:TfreeVal}), and thus the total time is increased by a factor of $\sum_{\ell\geq0} 2^{-\ell} = 2$.

This factor of only $2$ is one of the reasons we present our alternative binning strategy to the one in Ref.~\cite{Allen:2025hdx}; the corresponding factor for that work is roughly $25$, so, even accounting for our need to perform four Grover searches instead of two, the order-one numbers appear more favorable for our strategy. Note, however, that we are omitting many other small constants that should be calculated numerically (e.g. the necessary $N_{\rm QSP}$\footnote{Although our much larger transition region---$\Phi \in (2\pi/15, \pi/5)$, i.e. $\Delta = \pi/15$, compared to a region of width $\Delta = \pi/380$ in Ref.~\cite{Allen:2025hdx}---means that our $N_{\rm QSP}$ is likely also much smaller.}), so this should be understood as purely suggestive, and not a definitive advantage; as we have emphasized, there are many possible binning strategies and we leave proper optimization over them to future work.

\begin{figure*}
    \centering
    \includegraphics[width=0.95\linewidth]{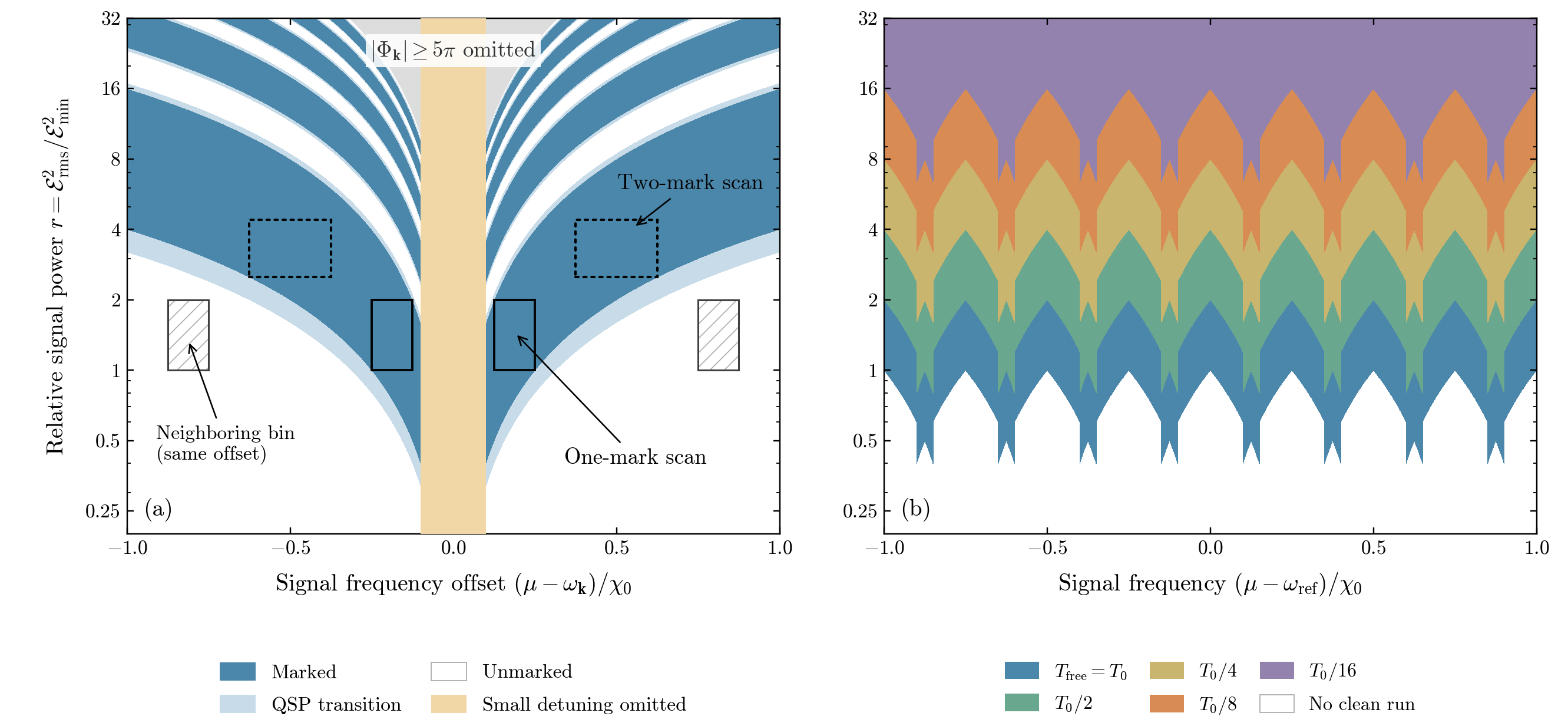}
    \caption{Illustrative phase responses and signal coverage for the four-offset binning scheme. The phase associated with register entry $\bk$ is $\Phi_\bk = r(\chi_0/\delta_\bk)(T_\free/T_0)(\pi/20)$, where $r = \sigErms^2/\sigEmin^2$ and $T_0 = \pi\chi_0/(20\sigEmin^2)$. We use idealized QSP windows that mark entries with $\pi/5 \leq \bar\Phi_\bk \leq 4\pi/5$, where $\bar\Phi_\bk = \Phi_\bk \pmod{\pi}$ lies in $[0,\pi)$.
    (a) Response of a single register bin with free evolution time $T_\free = T_0$. Blue indicates phases in the QSP acceptance window, while pale blue shows the transition regions; note that the shading uses the narrower transitions of width $\pi/25$ introduced for the two-mark example. The yellow region excludes small detunings, $|\delta_\bk|<\chi_0/10$, where adiabaticity may fail; its width is illustrative and should be evaluated numerically for a particular detector and error tolerance. Higher phase wraps with $|\Phi_\bk|\geq5\pi$ are omitted in gray. The solid black rectangles show the conservative one-mark regions, $\chi_0/8\leq|\delta_\bk|\leq\chi_0/4$ and $1\leq r\leq2$, which also work with the wider one-mark QSP transitions used in the main text. The hatched rectangles correspond to acceptance regions of the next bins over on either side, which are fully in the unmarked regime. The dotted rectangles show the two-mark regions discussed in the text, $3\chi_0/8\leq|\delta_\bk|\leq5\chi_0/8$ and $5/2\leq r\leq22/5$. 
    (b) Combined coverage using only one-mark searches, with all four offsets $\omof=0,\chi_0/4,\chi_0/2,3\chi_0/4$ and five free evolution times: $T_\free=T_0,T_0/2,\ldots,T_0/16$. Here $\omega_{\rm ref}$ is a fixed resonance of the unshifted comb. We use the main text's one-mark QSP, with transition widths $\pi/15$. A run contributes only if exactly one entry is marked and all other entries lie in the rejection windows; any entry in a transition region or within the small-detuning cutoff disqualifies that run. Blue, green, gold, orange, and purple identify coverage at the five durations, respectively. Where several runs qualify, we show the shortest free evolution time. White indicates that none of these runs qualifies. As prescribed in the main text, each successive duration covers the next factor of two in $r$, giving continuous frequency coverage throughout $1\leq r\leq32$. Additional response below $r=1$ is available at some frequencies.}
    \label{fig:PhaseCoverage}
\end{figure*}

In fact, the search procedure described above uses only a small portion of the detector’s response in the frequency–coupling plane. Our binning prescription ensures coverage for $1\leq r\leq2$, with the signal assigned to a register entry whose detuning satisfies $\chi_0/8\leq|\delta_\bk|\leq\chi_0/4$. However, QSP-accepted phases also arise outside these ranges. Writing the signal's fractional location between adjacent cavity resonance frequencies as $x = (\mu-\omega_\bk)/\chi_0$, our chosen interrogation time gives $|\Phi_\bk| \approx \pi r/(20|x|)$. The marked phase interval used above, $\pi/5 \leq |\Phi_\bk| \leq 4\pi/5$, therefore corresponds to $4|x| \lesssim r \lesssim 16|x|$. Further marked intervals recur as the accumulated phase times two (since the ancilla's relative phase is $2\Phi_\bk$, not $\Phi_\bk$) wraps around the unit circle at larger values of $r$, e.g. when $|2\Phi_\bk| = 3\pi$. As illustrated in Fig.~\ref{fig:PhaseCoverage}(a), the combined region of parameter space that will result in a $-1$ phase after the QSP includes both larger detunings at higher signal strengths, and $r<1$ at smaller detunings: although we rely only on signals in the two black rectangles applying the $-1$ phase, the full blue region that will give such a phase is much larger. Our construction is thus highly conservative, and a real experiment should likely include signals from these much larger areas.

Exploiting this additional response would require modifying both the Grover search and the subsequent candidate verification. At larger $r$ values, multiple register entries will typically acquire marked phases. As a concrete example of this, suppose that we expanded our QSP's rejection windows to $0 < \bar{\Phi} < 4\pi/25$ and $21\pi/25 < \bar{\Phi} < \pi$. Now note that, among the four $\omof$ values, there is a run where the nearest bin has a detuning
\begin{align}
    \left| \delta_{\bk'} \right| \in \left[ \frac{3\chi_0}{8}, \, \frac{\chi_0}{2} \right]
\end{align}
while the second-nearest bin has a detuning
\begin{align}
    \left| \delta_{\bk''} \right| \in \left[ \frac{\chi_0}{2}, \, \frac{5\chi_0}{8} \right].
\end{align}
Every other bin has a detuning $|\delta| \geq 11\chi_0/8$. We therefore have
\begin{subequations}
\begin{align}
    \left| \Phi_{\bk',\bk''} \right| &\in \left[ \frac{2\pi}{25}r, \, \frac{2\pi}{15}r \right] \\
    \left| \Phi_{\bk \neq \bk',\bk''} \right| &\leq \frac{2\pi}{55}r.
\end{align}
\end{subequations}
We will thus have both bins in the accepted region, and all others in the rejected region, over the range
\begin{align}
    \frac{5}{2} \leq r \leq \frac{22}{5}.
\end{align}
We could therefore perform a Grover search that looks for two marked bins---this amounts simply to reducing $N_{\rm Grover}$ to $\pi\sqrt{\Nbins/2}/4$---to search this range of larger $r$ values and larger detunings. We illustrate this additional range of frequencies and signal strengths with the dotted rectangles in Fig.~\ref{fig:PhaseCoverage}(a). The resulting Grover search is faster by a factor of $\sqrt{2}$, but randomly returns either one of the marked bins, so each must be checked for a signal. Adding this to our existing search is not immediately beneficial, as the $2 < r < 5/2$ window remains open and thus we would need to perform some additional search, but this illustrates the general point that a variety of search strategies, including ones involving multiple marked bins, can be adapted to dark matter detection.

At even larger $r$ values, several register entries will typically acquire marked phases, and these bins can become increasingly distant from the signal frequency as $r$ grows. A generalized Grover's algorithm capable of handling an unknown number of marked bins is known~\cite{Boyer:1996zf}, however, and the subsequent search of nearby frequencies would be unlikely to meaningfully affect total operation time: a sequential search of up to $O(\sqrt{\Nbins})$ bins around the output would have only an order-one impact on the total operation time even if one searched all the way to $\sigEmin$ in each bin, and in fact this is greatly excessive: small signals could only output a bin close to $\mu$, and thus more distant bins need to be searched only down to larger couplings. Moreover strong signals would likely appear in several different Grover searches with different parameters (e.g. different $\sigEmin$ values); combining the information from these Grover searches before performing any single-frequency follow-ups would likely improve efficiency further.

An additional complication is that, with a finite QSP sequence, bins may end up in a phase-digitization transition region, where the register and ancilla develop problematic entanglement, as one attempts to include a large range of $r$ values. A reliable extension of the protocol must therefore account for this transitional behavior as well.

Nonetheless, these observations suggest optimizing the search by arranging overlapping coverage of the frequency–coupling plane. Frequency offsets translate the response region (i.e. the blue region in Fig.~\ref{fig:PhaseCoverage}(a)) horizontally, while changing the free evolution time shifts them vertically. Figure~\ref{fig:PhaseCoverage}(b) illustrates how our usual four frequency combs ($4\omof/\chi_0=0,\,1,\,2,\,3$) with five values of $T_\free$ can tile a region of the frequency-coupling plane with at least one bin appearing at an appropriate distance from every signal frequency and strength. More general strategies could exploit distant marked entries, adjust the bin spacing or digitization thresholds, and choose subsequent measurements adaptively, in order to achieve reliable signal detection across the desired parameter range with minimum total observation time, including the costs of phase digitization, repeated searches, and candidate verification. We leave this optimization to future work; the construction presented here is sufficient to establish a parametric advantage without making full use of the detector’s response.

\section{Technical Requirements}\label{sec:TechReqs}

The discussion up to this point has, with a few exceptions, assumed perfect operation of the detector and limiting behaviors of some of its steps, with errors and deviations from those limits omitted almost everywhere. In this section we discuss several technological requirements necessary to make this a good approximation. The full list of issues that we consider is as follows:
\begin{itemize}
    \item Errors (including decays, excitations, and phase errors) of the register qubits. This is the most serious class of errors for our proposal, and is the main reason we do not believe that a detector with the exact form of our toy model could be built in the near future. Below we discuss both the severity of this problem, and some potential mitigating strategies.
    \item Decays (or excitations) of the cavity. Our Grover implementation requires coherent superpositions of the cavity’s $\ket{0_c}$ and $\ket{1_c}$ states, making photon loss and cavity dephasing important throughout the search. The cavity coherence time must therefore exceed the full Grover interrogation time. This is a substantial requirement, although potentially less restrictive than the requirement on the qubit register above. Notably, the coherent phase accumulator alone is much less sensitive to photon loss, because its cavity occupation is very small throughout the search.
    \item Imperfect adiabaticity (both during the free evolution and during the signal ramps on and off). As we discussed previously, limiting this sets the minimum detuning and $T_\ramp$ that must be used in order for our phase estimate to be accurate. Additionally, imperfect adiabaticity may lead to residual cavity occupation after the signal ramp down; this is a more serious concern, but we show below that it becomes inconsequential with appropriate bin widths.
    \item Imperfect QSP behavior. This has already been discussed: the finite error $\xi$ in the $P = \pm1$ regions has only a logarithmic impact on our total runtime.
    \item The stochasticity of dark matter. In part, we have already discussed this: our choice of larger detunings compared to Ref.~\cite{Allen:2025hdx} ensures that time-variation in dark matter's phase does not affect us. There is an additional subtlety from signal amplitude variation, which we discuss briefly below.
    \item Calibration (i.e. bin frequency) errors. Deviations from our assumption of equally-spaced bins can modify the detector behavior somewhat; we discuss this below.
    \item Counter-rotating terms. Although we kept only co-rotating terms in our discussion above, this is not a good approximation at very large detunings. We show below that this does not affect our conclusions meaningfully. 
    \item Imperfect decoupling. A signal-decoupling scheme, such as the dynamical decoupling scheme we describe, may not be perfect; we briefly consider its imperfections later in this section.
    \item Gate errors. Although we leave a full study of our algorithm in the presence of imperfect gates to future work, we perform a rough, conservative estimate of the gate error rates below.
\end{itemize}

Other than what is needed to handle these errors, our toy detector would require a few additional components. First, we have assumed that it is possible to achieve $\chi_k$ values large enough to shift $\omega_c$ by an appreciable fraction of unity, since the range of dark matter masses our experiment could search for is
\begin{align}
    \frac{\mu_{\rm max}}{\mu_{\rm min}} \approx \frac{\max_\bk (\omega_\bk)}{\omega_c} \approx 1 + \frac{2\chi_{n_q-1}}{\omega_c}.
\end{align}
For example, a qubit-conditioned resonator frequency shift of $\chi/(2\pi)=366\,\mathrm{MHz}$ has been demonstrated using a quarton coupler and a nearly linearized transmon mode at approximately $5.1\,\mathrm{GHz}$~\cite{Ye2025}. If this coupling scale could be realized for the largest $\chi_k$ in our detector, it would give a mass range of order $10\%$. Achieving such a coupling together with the cavity and register coherence required here remains a separate challenge. For classical tuning of a three-dimensional cavity, Ref.~\cite{Zhao2025} estimates a range of approximately $200\,\mathrm{MHz}$ by exploiting the avoided crossing with a flux-tunable SQUID mode.

This can be improved substantially further however, using two techniques: First, one can simply replace the larger-$\chi_k$ qubits with a collection of qubits, all at smaller $\chi_k'$ and all placed in the same state, such that their collective effect is equivalent to the original $\chi_k = 2^k \chi_0$. (Such a repetition code also helps with certain kinds of errors, as we discuss below.) Second, our strategy can likely be adapted to heterodyne detection (see e.g.~\cite{Berlin2020Conversion,Lasenby2020,Berlin2021Heterodyne}), in which the dark matter mass is much smaller than the cavity frequency; this allows even small fractional changes in the cavity frequency to potentially correspond to large changes in dark matter mass. The latter is a particularly attractive option because it can also increase the maximum benefit of a Grover enhancement: although our strategy is likely limited to $\Nbins \sim 10^5$ before one begins changing the cavity behavior by order-one, likely leading to large non-linear effects, a heterodyne detector could potentially have a substantially larger number of bins, and thus a much larger $\sqrt{\Nbins}$ enhancement.

Our Grover search additionally requires confinement to the cavity’s two lowest Fock states. A conservative sufficient condition for achieving this with a Kerr nonlinearity is an anharmonicity $|\kappa_{\rm Kerr}|$ much larger than the scanned bandwidth $W$: this keeps the unwanted transition between $\ket{1_c}$ and $\ket{2_c}$ far off resonance throughout the search, and thus suppresses its virtual contribution relative to the desired signal-induced shift. The additional phase per sensing interval from $|\kappa_{\rm Kerr}| < \infty$ is of order $\int dt\,|\sigE(t)|^2/|\kappa_{\rm Kerr}| \sim \chi_0/|\kappa_{\rm Kerr}|$; conservatively bounding its accumulation over $N_{\rm QSP} \sqrt{\Nbins}$ intervals to be small thus requires only $|\kappa_{\rm Kerr}| \gg W N_\QSP / \sqrt{\Nbins}$, likely weaker than the $|\kappa_{\rm Kerr}| \gg W$ assumption above for typical applications. A photon blockade must similarly suppress both real transitions and unwanted virtual energy shifts; we leave the precise conditions in this case to future work using specific strategies.

These conditions are deliberately stronger than may be necessary: Searches over suitably chosen, non-contiguous sets of frequency bins may be able to avoid the unwanted transition resonances without requiring $|\kappa_{\rm Kerr}| \gg W$, and incorporating predictable distortions into the QSP construction could potentially reduce the required phase accuracy. Such strategies could substantially relax the confinement requirements, although their coverage, additional scanning overhead, and accumulated errors require a separate analysis.

Another necessary component of our technique is an implementation of Grover's algorithm. Although the number of qubits on which this needs to be conducted is fairly small---even the $\Nbins \sim 10^5$ benchmark above corresponds to only $\approx\!17$ qubits---this is larger than anything that has been done to-date; $n_q = 5$ has been demonstrated~\cite{PokharelLidar2024}. Nonetheless we suspect that this is a smaller technical challenge than sufficiently decoupling higher Fock states and than some of the issues below.

\subsection{Register Qubit Errors}\label{sub:RegisterErrors}

Our application of Grover's algorithm requires coherence to be maintained in the register qubits from the moment the qubits are first prepared into the $\ket{+_k}$ states to the final measurement. Crucially, phase ($Z$) errors in these register qubits cannot be protected with quantum error correction, as they are fundamentally indistinguishable from a signal (since $e^{i\Phi Z} = \cos\Phi + i\sin\Phi Z$ and the constant component is of course unobservable). Other errors may or may not be possible to correct: $X$ errors should generally be correctable with an appropriately-chosen code, but decays $\sigma_-$ likely cannot be for our dispersive coupling, as correcting qubit decay will generally also eliminate our signal; see also the closely related Hamiltonian-Not-in-Lindblad-Span (HNLS) condition~\cite{Zhou2018HNLS}. The procedure we present would therefore require
\begin{align}
    \frac{n_Q\left( \Gamma_1 + \Gamma_\phi \right) N_{\rm QSP} \sqrt{\Nbins}}{\tau \sigEmin^2} \ll 1, \label{eq:RegisterErrorCondition}
\end{align}
where $\Gamma_1$ ($\Gamma_\phi$) is the error rate of decay (phase) errors. This is an extremely taxing requirement; at the smallest $\sigErms$ values, this means that there can be no register errors in an order-one fraction of the experiment's entire operation time---typically weeks to years in order to achieve competitive sensitivity. Although our Grover speedup may make it possible to achieve some advantage over shorter times, the necessary register qubit lifetime would nonetheless be far longer than any current physical superconducting qubit can achieve~\cite{Bland2025}.

As a result, we do not expect the exact experimental design we describe to be a viable path towards searching for dark matter; our goal, as we have emphasized, is to present a toy detector model demonstrating that the obstacles to Grover-enhanced dark matter searches are technological, not fundamental, but a great deal more work will be needed to make such a search feasible. Many directions should be explored in the attempt to resolve this; here we merely offer a few suggestions:
\begin{itemize}
    \item As we noted above, some errors in the register qubits can be corrected. Thus, for example, a repetition code (i.e. making each of the $n_q$ qubits actually be several qubits, all prepared in the same state and with the same coupling) allows one to correct $X$ errors with essentially no other changes to the system, since $Z$ remains transverse. To handle the more serious errors, however, one would need to engineer more involved cavity-qubit interactions (e.g. terms of the form $\hat{n}_k\hat{n}_\ell \hat{a}^\dagger \hat{a}$), such that the signal is no longer in the same direction as any combination of standard single-qubit errors (e.g. $Z$); we leave studying concrete implementations of this to future work.
    
    \item We assumed in Eqs.~\eqref{eq:Hqubit}-\eqref{eq:Hdisp} that both the qubit and dispersive couplings are proportional to $\hat{n}_k$, i.e. to $Z$. Although this is typical for transmon-cavity couplings, it may be possible to circumvent it; see, for example, related ideas in Refs.~\cite{Richer:2017ggb, Noh:2021gsf}. This could allow one to prepare the qubits initially in their ground state, but with that ground state already giving some superposition of cavity modes. This prevents both decays and phase errors in the initial state, while excitations are exponentially suppressed at temperature below $\omega_q$. Unfortunately, this does not fully solve the problem above, as qubits will become rotated away from the ground state by the signal; nonetheless this occurs only for $\ket{\bk^*}$ and, to a lesser extent, a few nearby values of $\ket{\bk}$, and thus it reduces the problem, so it may be one part of a solution.

    \item More broadly, there are straightforward analogues of our toy Hamiltonian, Eq.~\eqref{eq:Htotal}, in other physical systems. One may, for example, be able to use other physical modes of the cavity in place of the transmons. Superconducting radio frequency cavities have been demonstrated with many simultaneous modes with lifetimes far exceeding those available in transmons~\cite{Naik2017,Kim2025}.

    \item Similarly, our statement above that zero register errors must occur during the entire operation time is almost certainly too conservative. It is known that an advantage can be retained in Grover searches even when the oracle has a certain amount of noise~\cite{Vrana2014,RegevSchiff2008}, and while this does not parametrically change Eq.~\eqref{eq:RegisterErrorCondition}, it may improve order-one values and suggests that a reduced degree of coherent advantage may be possible even when Eq.~\eqref{eq:RegisterErrorCondition} is violated.
\end{itemize}
Nonetheless, none of these improvements are likely to be sufficient to salvage this work's specific design in the near term; more error-tolerant techniques will be necessary before a Grover-enhanced dark matter search can actually be implemented.

\subsection{Cavity Errors}\label{sub:CavityErrors}

We begin by considering cavity errors in the coherent phase accumulator alone, where they will be relatively minor, before turning to the full Grover search. Errors in the cavity state, like errors in the register qubits, can lead to decoherence between the $\ket{\bk}$ states. Unlike the register qubits, however, the cavity spends the entire operational time very close to vacuum, making errors far less likely. In particular, requiring that there are zero photon decays in the operation time is equivalent to the condition that
\begin{align}
\begin{split}
    \Gamma_\gamma \overline{\expect{a^\dagger a}} T_{\rm free}
    &\sim \Gamma_\gamma \sigEmin^2 \tau^2 \frac{1}{\sigEmin^2 \tau} \\
    &\sim \Gamma_\gamma \tau \ll 1,
\end{split}
\end{align}
where $\overline{\expect{a^\dagger a}}$ is the time-averaged expected cavity occupation and $\Gamma_\gamma$ is the single-photon decay rate. Rewriting, this is
\begin{align}
    \frac{Q_c}{Q_{\rm DM}} \gg 1,
\end{align}
where $Q_c = \omega_c/\Gamma_\gamma$ is the cavity quality and $Q_{\rm DM} \approx \omega_c \tau \approx 10^{6}$ is the quality of the dark matter signal. This is comfortably achievable by superconducting radio frequency (SRF) cavities, which have been built with qualities as high as $10^{10}$~\cite{Romanenko2020,Cervantes2024}. Integrated SRF systems with quantum control elements with $Q$ factors approaching $10^9$ have also been demonstrated~\cite{Pietikainen2024,Milul2023,Kim2025}. Noise in a coupled control qubit can also dephase the cavity;
for a cavity quantum memory, such dephasing has been experimentally
suppressed by monitoring the control qubit and applying real-time
feedback~\cite{Goldblatt2024}.

The phase accumulator likewise requires zero thermal excitations per observation time. This can be ensured simply by going to sufficiently low system temperature $T_{\rm sys}$, at which point a Boltzmann suppression $e^{-\omega_c/T_{\rm sys}}$ replaces the occupation number suppression $\overline{\expect{a^\dagger a}}$. We thus require
\begin{align}
    \Gamma_\gamma e^{-\omega_c/T_{\rm sys}} \frac{1}{\sigEmin^2 \tau} \ll 1,
\end{align}
or, rewriting,
\begin{align}
    T_{\rm sys} \ll \frac{\omega_c}{\ln\left( \frac{Q_{\rm DM}}{Q_c} \frac{1}{\sigEmin^2 \tau^2} \right)}.
\end{align}
This is certainly a taxing requirement---the right-hand side is of order 10 mK over a log factor for $\omega_c \sim \mu$eV---but it is not beyond the realm of possibility for a future detector; the CUORE experiment has previously demonstrated a macroscopic system cooled to 6.3~mK temperatures~\cite{Alduino2019}.

The demands for a Grover search are much more severe. The cavity used in our Grover implementation has an order-one occupation for nearly the entire Grover search time. Photon loss is consequently no longer suppressed by the small signal-induced occupation. Without an error-correction scheme, we conservatively require a negligible probability of photon loss over the entire search,
\begin{align}
    \Gamma_\gamma T_{\rm free} N_{\rm QSP}\sqrt{N_{\rm bins}} \ll 1,
\end{align}
Cavity dephasing must satisfy an analogous condition. Although demanding, this requirement is likely still less restrictive than that for the qubit register: as we noted above, cavity lifetimes can be longer, and there is no additional factor counting the register qubits. Note however that the relevant lifetime is that of the cavity with the Kerr or blockade mechanism operating, which need not retain the coherence of the uncoupled cavity.

\subsection{Imperfect Adiabaticity}\label{sub:ImperfectAdiabaticity}

In the main text, we assume that, subject to our conditions on binning and on ramping time, the cavity state follows its instantaneous extremal eigenvalue perfectly adiabatically. This will of course not be exactly correct, which leads to two deviations from the results in the main text: First, there will be small changes to the cavity's acquired phase; these should be evaluated numerically for a particular experimental proposal and may affect the optimal binning strategy, but they do not affect anything qualitatively. Second, and more importantly, imperfect adiabaticity can lead to leftover cavity occupation at the end of the ramp down, which could increase decay probabilities and lead to decoherence of the register qubits because they remain entangled with the cavity.

To estimate the latter effect, we integrate Eq.~\eqref{eq:dalphadt}, including the decoupling-induced factor of $\eta$. This gives an exact final state for the cavity for the coherent phase accumulator,
\begin{align}
    \alpha_f = -i \int_{t_i}^{t_f} dt \, \eta(t) \sigE(t) e^{-i\delta_\bk (t_f-t)}, \label{eq:FinalCavityAlpha}
\end{align}
corresponding to a probability of at least one photon left in the cavity of $P_{\geq 1} = 1 - e^{-|\alpha_f|^2}$. For a signal of zero linewidth (i.e. constant $\sigE$), and using linear ramps $\eta(t)$ of duration $T_\ramp$ each, this evaluates simply to
\begin{align}
    \left| \alpha_f \right| \sim \frac{\sigE}{\delta_\bk^2 T_\ramp}
\end{align}
and thus
\begin{align}
    P_{\geq 1} \sim \frac{\sigE^2}{\delta_\bk^4 T_\ramp^2}.
\end{align}
This should be made small enough to both avoid cavity decays (see Section~\ref{sub:CavityErrors}) and to have a negligible effect on the Grover success probability, even after summing for $N_\QSP\sqrt{\Nbins}$ iterations, but the very steep dependence on $\delta_\bk \sim \chi_0$ means this is unlikely to meaningfully affect our sensitivity.

For dark matter with a finite coherence time, however, the story is slightly more involved. Omitting the short ramp times, the integral in Eq.~\eqref{eq:FinalCavityAlpha} is, after averaging over possible dark matter states, the square root of the power spectral density $S(\delta_\bk)$ of dark matter at the given detuning, and thus
\begin{align}
    P_{\geq 1} \leq \expect{\hat{a}^\dagger \hat{a}}_f \sim T_\free S(\delta_\bk).
\end{align}
Dark matter's power spectral density is not known to high precision, but it has a fairly strict upper cutoff corresponding to the galactic escape velocity, which is only a few times the virial velocity~\cite{Gelmini:2020kcu, ParticleDataGroup:2024cfk}. Consequently, a detuning (i.e. $\chi_0$) of even a few times $\tau^{-1}$ should be sufficient to eliminate this contribution, up to small corrections from transient dark matter, substructure, etc.

For the Grover search, all of this will receive corrections from the suppressed higher Fock levels, which renders the adiabatic trajectories only approximate. Our estimate of the final cavity occupation is also no longer the right quantity to evaluate, since the cavity intentionally has an order-one occupation during the sensing time, but the same integral will set the unwanted transition amplitude between the two levels.

\subsection{Dark Matter Stochasticity}\label{sub:DarkPhotonStochasticity}

In writing Eq.~\eqref{eq:FreeEvolPhase}, we replaced the integral of $|\sigE(t)|^2$ by its average times $T_\free$. In the usual stochastic description of virialized wave-like dark matter, both the phase and amplitude of the field fluctuate by order one on the coherence timescale $\tau$. This applies to axions as well as dark photons; large occupation numbers do not suppress these amplitude fluctuations~\cite{Centers2021,Foster2018Halo}. For dark photons, variations in the polarization can also change the effective coupling to the cavity~\cite{Caputo2021}. These effects are all included in the effective drive $\sigE(t)$.

Our protocol depends on the time integral of the intensity $|\sigE(t)|^2$. Contributions to the dispersive phase therefore have a fixed sign, set by $\delta_\bk$, even as the dark matter phase changes. For intensity fluctuations with correlation time of order $\tau$~\cite{Derevianko2018Correlations,Foster2021Interferometry,Cheong2025WaveDM}, an exposure $T_\free\gg\tau$ averages over many approximately independent realizations, giving
\begin{subequations}
\begin{align}
    \overline{\Phi_\bk}
    &\simeq \frac{\sigErms^2T_\free}{\delta_\bk}, \\
    \frac{\sigma_{\Phi_\bk}}{|\overline{\Phi_\bk}|}
    &\sim \sqrt{\frac{\tau}{T_\free}}.
\end{align}
\end{subequations}
Here the overline denotes an average over realizations of the dark matter field, with $\sigErms^2=\overline{|\sigE(t)|^2}$, and $\sigma_{\Phi_\bk}$ is the standard deviation of $\Phi_\bk$ for individual free evolution intervals. Thus order-one fluctuations in the instantaneous signal produce only a small fractional fluctuation in the accumulated phase. For an order-one target phase and fixed $|\delta_\bk|\tau$, this fractional fluctuation scales as $\sigErms\tau$, which is small in the weak-signal regime considered here.

The remaining fluctuations must also be small over the full quantum computation. Successive long interrogations have approximately independent fluctuations, so, for order-one input phases, their leading root-mean-square effect over $N_\QSP$ queries scales as $\sqrt{N_\QSP\tau/T_\free}$. A conservative sufficient condition is
\begin{align}
    \sqrt{\frac{N_\QSP\tau}{T_\free}}
    \ll \frac{1}{\sqrt{\Nbins}}.
\end{align}
Subject to this accuracy requirement, stochastic amplitude fluctuations do not alter the parametric sensitivity or prevent coherent phase accumulation.

\subsection{Imperfect Calibration/Binning}\label{sub:BinningErrors}

Throughout the text, we have assumed that the frequency comb established by the register qubits is exactly evenly spaced, with $\chi_0$ between every pair of bins $\bk$ and $\bk+1$. This may be difficult to do in practice, considering that the $\chi_k$ values span a range of a factor of $2^{n_q-1}$. We can therefore consider errors in which the actual central frequency associated with $\ket{\bk}$ is $\omega_c + \omof + \chi_0\bk + \delta\omega_\bk$, rather than merely the first three terms.

For measurable, time-independent bin frequency errors $\delta\omega_\bk$, requiring that all $\delta\omega_\bk \ll \chi_0$ should be essentially sufficient: while one would likely want to slightly adjust bin assignments, $T_{\rm free}$, etc. in order to accommodate the slightly different phases acquired by bins with a non-zero $\delta\omega_\bk$, this should not meaningfully complicate the procedure. If these errors cannot be measured, however, they become more problematic, as they lead to uncorrectable phases $\delta \nu_\bk = \int dt \, \delta\omega_\bk(t)$ per free evolution interval; a conservative sufficient requirement is then that, for all $\bk$,
\begin{align}
    \delta\nu_\bk \ll \frac{1}{N_\QSP \sqrt{\Nbins}},
\end{align}
since the phase error can, at worst, add linearly in the number of repetitions. Time-dependent frequency errors in particular are likely to fall into this unmeasurable category, unless one has a means of measuring them in real time. We leave detailed study of this, including numerical simulation of QSP behavior in the presence of phase instability, to future work on specific detector proposals.

\subsection{Counter-Rotating Terms}\label{sub:CounterRotatingTerms}

In our calculation of the behavior of the cavity during its free evolution, we kept only co-rotating terms in order to simplify the discussion. In fact, this is not necessary: the cavity behavior is exactly calculable with the full dark matter signal (for the coherent phase accumulator; as elsewhere in this section, the trajectory receives corrections in the Grover search). We briefly explain this here in order to illustrate the impact of counter-rotating terms on our protocol. For simplicity we omit the finite coherence time of dark matter here, treating it as a pure trigonometric function, but the generalization to arbitrary Fourier spectra is apparent.

Including the counter-rotating terms, Eq.~\eqref{eq:effHcav} becomes
\begin{align}
    H_{\rm eff} = \delta_\bk \hat{a}^\dagger \hat{a} + \epsilon (a^\dagger + a) + \epsilon (a^\dagger e^{2i\mu t} + a e^{-2i\mu t}),
\end{align}
where we have absorbed the (constant) phase $\phi$ into the definition of $a$ and $a^\dagger$. This has an exact solution:
\begin{align}
    \alpha(t) = -\epsilon \left( \frac{1}{\delta_\bk} + \frac{e^{2i\mu t}}{\Sigma_\bk} \right)
\end{align}
where $\Sigma_\bk = \mu + \omega_\bk$. The phase accumulated over times $T_{\rm free} \gg 1/\delta_\bk$ is then
\begin{align}
    \Phi_\bk \approx \epsilon^2 T_{\rm free} \left( \frac{1}{\delta_\bk} + \frac{1}{\Sigma_\bk} \right).
\end{align}
In the main text, we omitted the second term, but for $\delta_\bk \sim \mu$ this is no longer a good approximation. Nonetheless this correction is small for any of the bins where $\Phi_\bk$ is large enough to be relevant, unless $r$ is extremely large (far above our cautious limit of $r<2$), so this does not meaningfully change our conclusions. This does, however, affect the shape of the ``wings'' in the left plot of Fig.~\ref{fig:PhaseCoverage}, although only at signal frequency offsets of order $\mu$, far outside of that plot's range.

\subsection{Imperfect Decoupling}\label{sub:ImperfectDecoupling}

\begin{table*}[t]
    \renewcommand{\arraystretch}{3}
    \begin{tabular}{c @{\hspace{5pt}}|@{\hspace{5pt}} c @{\hspace{5pt}}|@{\hspace{5pt}} c @{\hspace{5pt}}|@{\hspace{5pt}} c}
    \multicolumn{4}{c}{\textbf{Scaling Comparison}} \\
    \hline
    \textbf{Strategy} & $\displaystyle \sigEmin(W,T)$ & $\displaystyle T(\sigEmin, W)$ & $\displaystyle W(\sigEmin, T)$ \\
    \hline
    Sequential Search & $\displaystyle \frac{W^{1/2}}{T^{1/2}}$ & $\displaystyle \frac{W}{\sigEmin^2}$ & $\displaystyle \sigEmin^2 T$ \\
    \makecell{Grover-Heisenberg Search\\($\tau\to\infty$)} & $\displaystyle \frac{W^{1/3}}{T^{2/3}}$ & $\displaystyle \frac{W^{1/2}}{\sigEmin^{3/2}}$ & $\displaystyle \sigEmin^3 T^2$ \\
    This Work & $\displaystyle \frac{W^{1/4}}{T^{1/2}\tau^{1/4}}$ & $\displaystyle \frac{W^{1/2}}{\sigEmin^2 \tau^{1/2}}$ & $\displaystyle \sigEmin^4 T^2 \tau$
    \end{tabular}
    \caption{A parametric comparison of an ideal sequential search, the original Grover–Heisenberg search for an indefinitely coherent signal, and the finite-coherence protocol developed in this work. Here $\sigEmin$ is the minimum detectable drive amplitude, $W \sim \omega_c \Nbins/10^6$ is the frequency bandwidth searched, $T$ is the total observation time, and $\tau$ is the (dark matter) signal coherence time, $\tau \sim 10^6/\mu$. The three columns express the same sensitivity relation keeping different resources fixed, with numerical constants and logarithmic overheads omitted. The last row, describing this work, applies in the weak-signal, many-bin regime $\sigEmin\tau \ll 1$ and $W\tau \gg 1$.}
    \label{tab:ScalingComparison}
\end{table*}

In the main text, we noted that an approach to decoupling the signal from the cavity is necessary in order to adiabatically follow the dark matter-induced instantaneous extremum of the cavity Hamiltonian in the co-rotating frame, and we suggested dynamical decoupling as a means of doing so. As discussed in Section~\ref{sub:Ramping}, in a dynamical decoupling scheme one employs rapid $\Pi = e^{i\pi a^\dagger a}$ operations, which effectively flip the sign of $\sigE(t)$. For complete decoupling, one can apply a $\Pi$ every time $\Delta t \ll 1/\delta_\bk$, with the duration $T_\DD$ of each $\Pi$ operation assumed to be much shorter than $\Delta t$. For partial decoupling, $0 < \eta < 1$, one can instead use an asymmetric square wave, with alternating intervals of $\Delta t_+$ and $\Delta t_-$ between the $\Pi$ operations; this results in
\begin{align}
    \eta = \left|\frac{\Delta t_+ - \Delta t_-}{\Delta t_+ + \Delta t_-}\right|.
\end{align}

Outside of the idealized dynamical decoupling limit, a few corrections arise. First, even for $\Delta t_+ = \Delta t_-$, quadratic effects of the dark matter signal will persist; these lead to a residual phase accumulation
\begin{align}
    \left.\frac{d\Phi}{dt}\right|_{\eta = 0} \sim (\delta_\bk \Delta t)^2 \left.\frac{d\Phi}{dt}\right|_{\eta = 1}.
\end{align}
This will lead to small phase corrections that will depend on the time spent outside of free evolution; these are unlikely to be more than small corrections but should be evaluated for any specific implementation of our design.

Second, since we must actually ramp the $\eta$ between $0$ and $1$ over a time $T_\ramp$, the Fourier spectrum of $\eta(t)$ will in fact be supported at non-zero frequencies of order $1/T_\ramp$. If these become comparable to $\delta_\bk$, the effective signal seen by the cavity could come on or close to resonance, resulting in potentially large erroneous phases and/or significant occupation of the cavity. So long as one uses $T_\ramp \gg 1/\delta_\bk$, however, this should not be an issue.

Third, and relatedly, the implementation of $\eta \neq 0$ could additionally create problematic harmonics, even in the $T_\ramp \to \infty$ limit, if not handled correctly: the Fourier spectrum of the square wave described above is supported at frequencies
\begin{align}
    \Omega_m = \frac{2\pi}{\Delta t_+ + \Delta t_-}m
\end{align}
for integer $m$, which could reach frequencies as small as $\delta_\bk$ if $\Delta t_+ + \Delta t_-$ became sufficiently large, for example if one attempted to take $\eta \to 1$ by increasing $\Delta t_+$ at fixed $\Delta t_-$. This is avoidable by instead changing the ratio of $\Delta t_+$ to $\Delta t_-$ with their sum kept fixed, although in this case one is ultimately limited by $\Delta t_\pm \gtrsim T_\DD$. This likely necessitates the need for a small discontinuity as $\eta \to 1$. Although this will lead to slight non-adiabaticity, its effect is essentially just to cause small oscillations in the value of $\alpha(t)$ (see Appendix~\ref{app:CavityEvolution}), which make little difference. A resulting small leftover cavity occupation at the end of the signal ramp off and the final CX gate can be safely eliminated by measuring the cavity state; in the unlikely event that a photon is seen, the run can be restarted.

A more problematic effect arises if one instead uses a decoupling strategy based simply on a large detuning, which one begins at a large value (such that all bins are far off resonance) and then continuously ramps to zero: in this case one may pass through zero detuning for some bins; this may cause problems, but we do not explore it further here.

\subsection{Imperfect Gates}\label{sub:GateErrors}

Finally, errors in the gates used for the QSP and for Grover's algorithm could of course lead to errors in the final results. The total number of gates needed is
\begin{align}
    N_{\rm gates} \sim N_{\rm QSP} \sqrt{\Nbins} n_q,
\end{align}
with the largest number coming from the known-phase corrections of each qubit, and thus a conservative condition of zero errors per run reduces to requiring that the gate error probability $p_g$ satisfy
\begin{align}
    p_g \lesssim \frac{1}{N_{\rm QSP} \sqrt{\Nbins} n_q},
\end{align}
omitting the distinction between different types of gates for simplicity. For reasonable values of $n_q$ and $N_{\rm QSP}$, this suggests $p_g \lesssim 10^{-4}$--$10^{-6}$, which is demanding but likely a lesser issue than the register qubit demands above. We leave a more thorough accounting of this---including different gate types, limited-connectivity effects, systematic gate errors that can add coherently but also potentially be calibrated away, etc.---to future treatments of specific detector proposals.

\section{Conclusion}\label{sec:Conclusion}

In this work, we have presented a protocol that uses Grover’s algorithm to accelerate a search for wave-like dark matter, even when detecting the signal requires integration over many dark matter coherence times. The essential ingredients are a dispersively-coupled qubit register that allows the cavity to be placed in a superposition of resonance frequencies, and an energy shift that is independent of the dark matter field's phase. By adiabatically controlling its effective interaction with dark matter, a cavity can accumulate a dark matter-induced phase and then return to the vacuum, leaving that phase encoded in the qubit register.  Quantum signal processing then converts this response into an approximate Grover oracle, after which one can apply Grover's algorithm to find a candidate signal frequency. The QSP operation may require a restriction of the cavity to just the first two levels, for example using a photon blockade~\cite{Chakram:2020eky} or Kerr non-linearity~\cite{Kirchmair:2012rsm}. This construction allows dark matter to participate coherently in quantum computation, despite its finite coherence time.

Table~\ref{tab:ScalingComparison} compares the parametric sensitivity, observation time, and accessible bandwidth of our protocol with those of an ideal sequential search and of the original Grover–Heisenberg protocol, which assumed an infinite (or effectively infinite) signal coherence time. Here, $W$ is the frequency bandwidth searched, $T$ is the total observation time, and $\sigEmin$ is the minimum detectable drive amplitude, which is proportional to the dark matter coupling. We omit numerical constants and logarithmic overheads. A sequential search requires time $T \sim W/\sigEmin^2$, whereas our protocol requires only $T\sim W^{1/2}/(\sigEmin^2\tau^{1/2})$. Thus, at fixed target sensitivity, the search is accelerated by a factor $\sim\sqrt{W\tau}$, i.e. by the square root of the number of frequency bins, each of width $\sim\tau^{-1}$; this is simply the familiar Grover enhancement for finding an item in a list of length $\Nbins$. Equivalently, for a fixed bandwidth and observation time, the minimum detectable coupling is reduced by a factor $\sim(W\tau)^{1/4}$ compared to a traditional sequential scan.

The finite coherence time of dark matter nevertheless has a cost: For a perfectly coherent signal, the original Grover–Heisenberg protocol achieves $T\sim W^{1/2}/\sigEmin^{3/2}$, giving $\sigEmin\propto T^{-2/3}$ at fixed bandwidth. Our protocol instead gives $\sigEmin\propto T^{-1/2}$, due to necessarily coarser frequency resolution. For an indefinitely coherent signal, the characteristic detuning and bin width can decrease with the target amplitude, giving an order-one phase in a time proportional to $\mathcal E_{\min}^{-1}$. In our finite-coherence construction, these scales depend on the dark matter linewidth, giving a free evolution time proportional to $(\sigEmin^{2} \tau)^{-1}$. In the weak-signal regime $\sigEmin \tau \ll 1$ relevant to dark matter detection, the broader bins that result reduce the number of Grover iterations at a fixed bandwidth, yielding a total search time $T \sim W^{1/2} / (\sigEmin^{2} \tau^{1/2})$.

Realizing this advantage experimentally remains extremely difficult. In particular, at least in our toy model, the register must preserve coherence throughout the entire quantum search, requiring qubit error rates far below those currently available. The protocol should therefore be understood as a proof of principle and a starting point for more practical detector designs. Important directions include developing interactions compatible with error correction, exploring alternative register encodings and potentially multimode cavities~\cite{Kim2025,Paik2016}, and numerically optimizing the frequency binning and phase digitization. More broadly, the dispersive phase accumulation developed here may enable other forms of coherent processing of dark matter signals. Finite field coherence does not itself prevent such processing, leaving quantum computation as a potential resource for future dark matter searches.

\acknowledgments

We thank Prateek Agrawal, Hannah Banks, Itay Bloch, Hajime Fukada, Anson Hook, Andrew Jordan, Yao Lu, Nick Rodd,  Tanay Roy, and Tanner Trickle for useful discussions. We also thank Yunjia Bao, Yu Cheng, Liang Jiang, and Lian-Tao Wang, for coordinating the release of this work concurrently with their work on the subject. This work was supported by the Office of High Energy Physics of the U.S. Department of Energy QuantISED program and  by the Department of Energy, Office of Science, National Quantum Information Science Research Centers, Superconducting Quantum Materials and Systems Center (SQMS) under contract number  89243024CSC000002. This manuscript has been authored by Fermi Forward Discovery Group, LLC under Contract No. 89243024CSC000002 with the U.S. Department of Energy, Office of Science, Office of High Energy Physics. This work was performed in part at the Aspen Center for Physics, which is supported by National Science Foundation grant PHY-2210452. 

Several versions of GPT-5.x/6.x, Opus-5.x, and Fable-5.x were used to assist in figure generation, performing preliminary calculations, and finding references. However all final calculations were performed by the authors, and all text was written by the authors, up to LLM-assisted proofreading.

\onecolumngrid
\makeatletter
\onecolumn@grid@setup
\let\set@footnotewidth\set@footnotewidth@one
\let\compose@footnotes\compose@footnotes@one
\makeatother

\appendix

\section{Resonant and Dispersive Cavity Evolution and Coherence Properties}
\label{app:CavityEvolution}

In this appendix we compare resonant photon counting with dispersive phase readout in the noiseless limit. To use our signal in Grover's algorithm, we need the cavity to return to vacuum while the register retains the accumulated phases, as in Eq.~\eqref{eq:NoisyOracleEvolution}. Adiabatic phase accumulation achieves this, allowing for coherent quantum processing. Here we focus more on our coherent phase accumulation as a sensing technique in its own right, and compare it to the more standard resonant displacement sensing. We will show that in the noiseless limit, these two techniques scale parametrically in the same way, and are thus interestingly complementary. Dispersive readout can thus have a parametric advantage only if further quantum processing is employed.

We first review the cavity dynamics underlying this operation, then compare its sensitivity and coherence properties with those of a resonant measurement. We neglect cavity decay, finite temperature effects, and other errors throughout this appendix.

\subsection{Rotating-Frame Evolution}
\label{subapp:RotatingEvolution}

As in the main text, we work in the frame co-rotating with the signal at frequency $\mu$. Conditional on the register being in $\ket{\bk}$, the cavity Hamiltonian of Eq.~\eqref{eq:HcavDD} is
\begin{align}
    H_{\rm rot}(t) = \delta_{\bk} \hat a^\dagger \hat a
    + \eta(t)\left(\sigE(t)\hat a^\dagger+\sigE(t)^*\hat a\right),
    \qquad \delta_{\bk}=\omega_{\bk}-\mu,
    \label{eq:appCavityH}
\end{align}
where $\sigE(t)=\epsilon(t)e^{-i\phi(t)}$ is the complex DM-induced drive. The control envelope $\eta(t)\in[0,1]$ varies the effective coupling, from decoupled at $\eta=0$ to fully on at $\eta=1$. For a given signal realization, a cavity initially in vacuum remains in a coherent state,
\begin{align}
    \ket{\psi_{\bk}(t)}=e^{i\Phi_{\bk}(t)}\ket{\alpha(t)},
    \qquad \alpha(0)=0,\quad \Phi_{\bk}(0)=0.
    \label{eq:appCavityState}
\end{align}
The dimensionless amplitude $\alpha=\langle\hat a\rangle$ specifies the cavity quadratures and gives the mean photon number $|\alpha|^2$. As in the main text, we leave its dependence on $\bk$ implicit. The amplitude and phase obey
\begin{align}
    \dot\alpha &= \underbrace{(-i\delta_{\bk}\alpha)}_{\text{rotation from detuning}}
    +\underbrace{(-i\eta\sigE)}_{\text{displacement from DM}},
    &\dot\Phi_{\bk} &= -\operatorname{Re}\!\left(\eta\sigE\alpha^*\right).
    \label{eq:appCavityEvolution}
\end{align}
It is useful to interpret the two terms in $\dot\alpha$ separately. The dark matter drive translates $\alpha$ in phase space with velocity $-i\eta\sigE$, independent of its position, while detuning rotates it about the origin with angular velocity $-\delta_{\bk}$. At resonance this rotation is absent, leaving only the displacement. To clarify the role of the various phases in the system, $\phi$ is the phase of the dark matter drive, while $\arg\alpha$ gives the direction of the cavity displacement. The factor $e^{i\Phi_{\bk}}$ changes the phase of the entire register component relative to the others, without changing that displacement.

\begin{figure}[tb]
    \centering
    \includegraphics[width=0.98\linewidth]{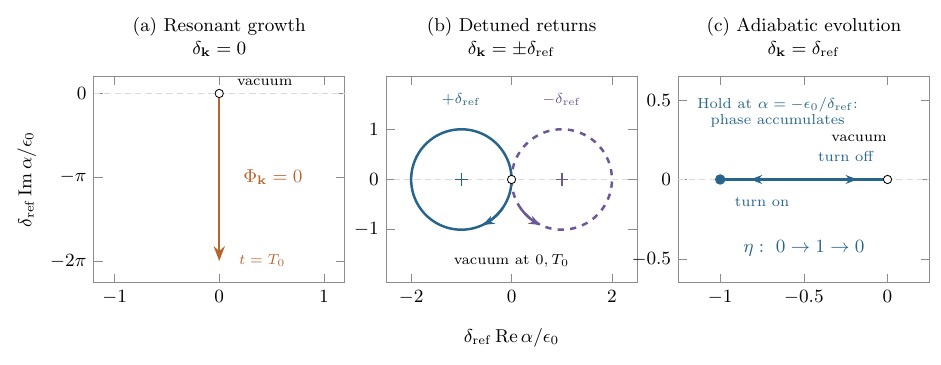}
    \caption{Cavity trajectories in the co-rotating frame for a fixed signal, with axes chosen so that $\sigE=\epsilon_0>0$ and a positive reference frequency $\delta_{\rm ref}$ setting the units. (a) Resonant displacement grows along a straight line, shown up to $T_0=2\pi/\delta_{\rm ref}$. (b) At $\delta_{\bk}=\pm\delta_{\rm ref}$, a constant interaction produces circles that return to vacuum at $T_0$. Crosses mark the points where DM translation and detuning-induced rotation cancel. (c) At $\delta_{\bk}=\delta_{\rm ref}$, adiabatic ramping carries the instantaneous eigenstate out and back along the same line. During the hold, its coherent amplitude is fixed while the state accumulates a phase. A finite ramp speed produces a small quadrature lag. The vertical scales differ between panels.}
    \label{fig:appTrajectories}
\end{figure}

We begin by holding both the dark matter amplitude and phase fixed, taking $\epsilon(t)=\epsilon_0$ and $\phi(t)=\phi_0$, so that $\sigE=\epsilon_0e^{-i\phi_0}$ is constant. The equations above also apply to a time-dependent signal; we will allow both $\epsilon$ and $\phi$ to vary in the next subsection. At resonance and with $\eta=1$, translation alone gives
\begin{align}
    \alpha(t)=-i\sigE t,
    \qquad \langle\hat a^\dagger\hat a\rangle=|\sigE|^2t^2,
    \qquad \Phi_{\bk}(t)=0\,,
    \label{eq:appResonantMotion}
\end{align}
as shown in the left panel of Figure~\ref{fig:appTrajectories}. Successive displacements (with the same phase) add coherently along a straight line. Changing $\phi_0$ rotates this line without changing the photon number. Turning off the interaction leaves the cavity displaced, so this simple resonant pulse does not implement Eq.~\eqref{eq:NoisyOracleEvolution}.

Away from resonance, with $\eta=1$, translation and rotation cancel at $\alpha=-\sigE/\delta_{\bk}$. Elsewhere their sum is $\dot\alpha=-i\delta_{\bk}(\alpha+\sigE/\delta_{\bk})$, perpendicular to the displacement from this point. The motion is therefore a circle about the shifted center. Starting from vacuum,
\begin{align}
    \alpha(t)=\frac{\sigE}{\delta_{\bk}}\left(e^{-i\delta_{\bk}t}-1\right).
    \label{eq:appDetunedMotion}
\end{align}
The circle has radius $|\sigE/\delta_{\bk}|$, and the sign of $\delta_{\bk}$ sets its direction of circulation.  At its center the two velocities cancel, giving a stationary state that we will discuss below. These circling states can still accumulate a phase. A circular trajectory returns to vacuum after $2\pi/|\delta_{\bk}|$, generally with a nonzero phase~\cite{Lemmer2013,Pechal2012}, but this return time differs between register branches. Figure~\ref{fig:appTrajectories} compares these trajectories with adiabatic evolution, which avoids the need to synchronize their returns, and is discussed below. Related dark matter sensing protocols use controlled displacements and qubit echoes~\cite{MaSheng2026}.

\subsection{Adiabatic Phase Accumulation}
\label{subapp:AdiabaticEvolution}

We can instead arrange for the cavity to adiabatically follow the instantaneous stationary state where rotation and displacement cancel each other, beginning and ending with vacuum. Completing the square in Eq.~\eqref{eq:appCavityH} gives its displacement and energy shift, as illustrated in Fig.~\ref{fig:shift} of the main text,
\begin{align}
    \alpha_{\rm ad}(t)=-\frac{\eta(t)\sigE(t)}{\delta_{\bk}},
    \qquad E_{0,\bk}(t)=-\frac{\eta(t)^2|\sigE(t)|^2}{\delta_{\bk}}.
    \label{eq:appDisplacedVacuum}
\end{align}
This eigenstate lies at the point where translation and rotation cancel. In the laboratory frame, this state is a driven oscillation at frequency $\mu$. For negative detuning it is a rotating-frame maximum rather than a minimum; the oscillation remains stable, and adiabatic following applies for either sign. As described in Section~\ref{sec:CoherentPhaseAcc} and illustrated in Fig.~\ref{fig:clock-branches} of the main text, adiabatic ramping of $\eta$ from zero to one and back lets the cavity follow this state and return to vacuum in every register branch. For a fixed signal, the limiting trajectory is the straight path shown in Fig.~\ref{fig:appTrajectories}(c).

We thus allow both $\epsilon(t)$ and $\phi(t)$ to fluctuate over the dark matter coherence time $\tau$~\cite{Gramolin2022,Centers2021}. Provided the cavity follows the displaced state to the required accuracy, the leading dispersive phase is
\begin{align}
    \Phi_{\bk}(T)\simeq-\int_0^T\!dt\,E_{0,\bk}(t)
    =\frac{1}{\delta_{\bk}}\int_0^T\!dt\,\eta(t)^2|\sigE(t)|^2.
    \label{eq:appFiniteCoherencePhase}
\end{align}
The sign follows from the $e^{i\Phi_{\bk}}$ convention in Eq.~\eqref{eq:appCavityState}. As discussed in Section~\ref{sec:CoherentPhaseAcc}, changes in the dark matter phase do not reverse this leading contribution, and intensity fluctuations change its rate but not its sign~\cite{MasiaRoig2023}. Ramping the interaction back off therefore returns the cavity to vacuum while leaving the phase on the register. As in Section~\ref{sec:CoherentPhaseAcc}, we require this return and the accumulated phase to be sufficiently accurate for every register component.

\subsection{Resonant and Dispersive Sensitivity}
\label{subapp:SensitivityComparison}

We now compare these two strategies, resonant displacement and adiabatic phase accumulation, at one chosen frequency. In this benchmark comparison we do not make use of the Grover search which is the main subject of this work. We assume a lossless cavity initially in vacuum, ideal register control and readout, and no background excitations. The measurement lasts for a time $T_\mathrm{free}\gg\tau$, with ramp times small compared with $T_\mathrm{free}$.

In a resonant search, the signal is stored in the cavity displacement. Within a coherence time its contributions add with approximately fixed phase, giving an occupation $n_{\rm res}\sim\sigErms^2T_\mathrm{ex}^2$ for an exposure time $T_\mathrm{ex}$. Over many coherence times they instead add as a random walk, giving~\cite{Shi2025,Freiman2026}
\begin{align}
    \overline{n_{\rm res}(T_\mathrm{tot})}\sim\sigErms^2\tau T_\mathrm{tot},
    \qquad \sigErms^2\equiv\overline{|\sigE(t)|^2},
    \label{eq:appResonantSearchSignal}
\end{align}
where the overline denotes an average over signal realizations. Each realization produces a coherent state, but its displacement direction depends on the changing dark matter phase. For a given displacement, the probability of detecting at least one photon is $1-e^{-|\alpha|^2}$. The unknown direction therefore does not prevent photon counting, and an occupation of order one gives an order-one detection probability~\cite{Lamoreaux2013,Dixit2021,Braggio2025}.

For dispersive readout, we prepare an equal superposition of a signal branch $\ket{\bk}$ and a sufficiently far-detuned reference branch $\ket{\mathbf l}$. After both return the cavity to vacuum, the state is
\begin{align}
    \ket{\psi(T)}=\frac{\ket{\mathbf l}+e^{i\vartheta(T)}\ket{\bk}}{\sqrt{2}}
    \otimes\ket{0_c},
    \qquad \vartheta=\Phi_{\bk}-\Phi_{\mathbf l}.
    \label{eq:appClockReadoutState}
\end{align}
Neglecting the reference phase and the short ramps, Eq.~\eqref{eq:appFiniteCoherencePhase} gives
\begin{align}
    \overline{\vartheta(T_\mathrm{free})} \simeq \frac{\sigErms^2T_\mathrm{free}}{\delta_{\bk}}.
    \label{eq:appClockSearchSignal}
\end{align}
We can read out this phase by interfering the two branches. In particular, projection onto $(\ket{\mathbf l}-\ket{\bk})/\sqrt{2}$ gives probability $\sin^2(\vartheta/2)$: this outcome never occurs in the absence of a signal, but has order-one probability once the accumulated phase is of order one. Although the energy shift is quadratic in the dark matter coupling, it is read out at the amplitude level through interference.

We note as an aside that this readout is reminiscent of an atomic clock, in which an atom is prepared in a superposition of two internal states. Their energy difference produces a relative phase, and a final pulse mixes the states so that their measured populations depend on that phase. Here the two components are register states, and dark matter produces the energy shifts we wish to measure.

Returning to our comparison, we hold the dimensionless detuning $|\delta_{\bk}|\tau\gtrsim1$ fixed, with the detuning and ramps chosen to achieve the accuracy needed for the oracle. The dispersive phase then scales as $\sigErms^2\tau T_\mathrm{free}$, up to a coefficient set by this dimensionless detuning and the line shape. The inverse-detuning approximation is controlled for $|\delta_{\bk}|\tau\gg1$ and gives a coefficient of magnitude $1/(|\delta_{\bk}|\tau)$; for detunings comparable to the linewidth, the phase coefficient and return error require the full signal spectrum. Setting the total time to equal the time of free evolution, and requiring either the resonant occupation or the dispersive phase to reach order one gives, at fixed $|\delta_{\bk}|\tau$,
\begin{align}
    \sigE_{\min}^{({\rm res})}\sim\frac{1}{\sqrt{\tau T_\mathrm{tot}}},
    \qquad
    \sigE_{\min}^{({\rm disp})}\sim\frac{1}{\sqrt{\tau T_\mathrm{tot}}}\,,
    \label{eq:appSensitivityScaling}
\end{align}
Both strategies therefore give the same parametric sensitivity to the dark matter coupling in this noiseless limit.

The difference is in their coherence properties. The resonant displacement depends on the changing dark matter phase, so contributions from different coherence intervals add as a random walk. The dispersive phase instead depends on the field intensity and accumulates with the same sign over many coherence times. After the cavity returns adiabatically to vacuum, this relative phase remains on the register and is available for further quantum processing, including Grover search.

\section{Quantum Signal Processing}\label{app:QSP}

In this appendix we explain the basic principle of quantum signal processing (QSP), elaborating on the brief introduction in the main text and describing a specific example. For illustration, we will actually use a slightly different procedure: generalized quantum signal processing (GQSP)~\cite{Motlagh:2023oqc}, which has the advantage that it can be applied directly to our coherent phase accumulator (using, for example, $n_q = 1$ with one frequency much less detuned than the other), without the need for the cavity modifications our Grover search demands.

Suppose that one has a single qubit and the ability to repeatedly apply an operation
\begin{align}
    U'_\Phi = 
    \begin{pmatrix}
        e^{i\Phi} & 0 \\
        0 & 1
    \end{pmatrix}
    =
    \begin{pmatrix}
        z & 0 \\
        0 & 1
    \end{pmatrix}
\end{align}
to it, where $\Phi$ is some unknown angle, as well as to apply arbitrary controlled rotations (by angles of one's choice) to the qubit. This is precisely the action of dark matter on our $n_q=1$ coherent phase accumulator when only one of two register states gets a significant phase. QSP allows us to construct, with $N_\QSP$ applications of $U'_\Phi$, any unitary of the form
\begin{align}
    U'_\QSP =
    \begin{pmatrix}
        P(z) & -z^{N_\QSP} Q(z)^* \\
        Q(z) & z^{N_\QSP} P(z)^*
    \end{pmatrix}
\end{align}
where $P$ and $Q$ are arbitrary polynomials of degree no greater than $N_\QSP$, satisfying $|P(z')|^2 + |Q(z')|^2 = 1$ for all $|z'| = 1$. In particular, by applying $U'_\QSP$ to $\ket{0}$, we can construct the transformation
\begin{align}
    U'_\QSP \ket{0} = P(z) \ket{0} + Q(z) \ket{1}.
\end{align}
The main text uses a different signal gate, with both phase signs, to approximate phases of $0$ or $\pi$ at different values of $\Phi$. Below we will describe a QSP which instead separates probabilities, approximating
\begin{align}
    |P(z)|^2 \approx 
    \begin{cases}
        0 \qquad |\tilde{\Phi}| < \frac{\pi}{2} \\
        1 \qquad |\tilde{\Phi}| > \frac{\pi}{2}
    \end{cases}
    \qquad\qquad\qquad|Q(z)|^2 \approx 
    \begin{cases}
        1 \qquad |\tilde{\Phi}| < \frac{\pi}{2} \\
        0 \qquad |\tilde{\Phi}| > \frac{\pi}{2},
    \end{cases}
\end{align}
where $\tilde{\Phi}$ is $\Phi \pmod{2\pi}$, with $\tilde{\Phi} \in [-\pi,\pi)$. We make this change because it works with this single-phase $U'_\Phi$, and because it allows the QSP operations to easily be visualized on the Bloch sphere; our main text QSP, which applies only a phase, is more difficult to visualize. Like the main-text QSP, approximations of these step functions will necessarily have transition regions of some width, within which the QSP will not give probabilities near 0 or 1.

\begin{figure}[t]
    \includegraphics[width=0.95\textwidth]{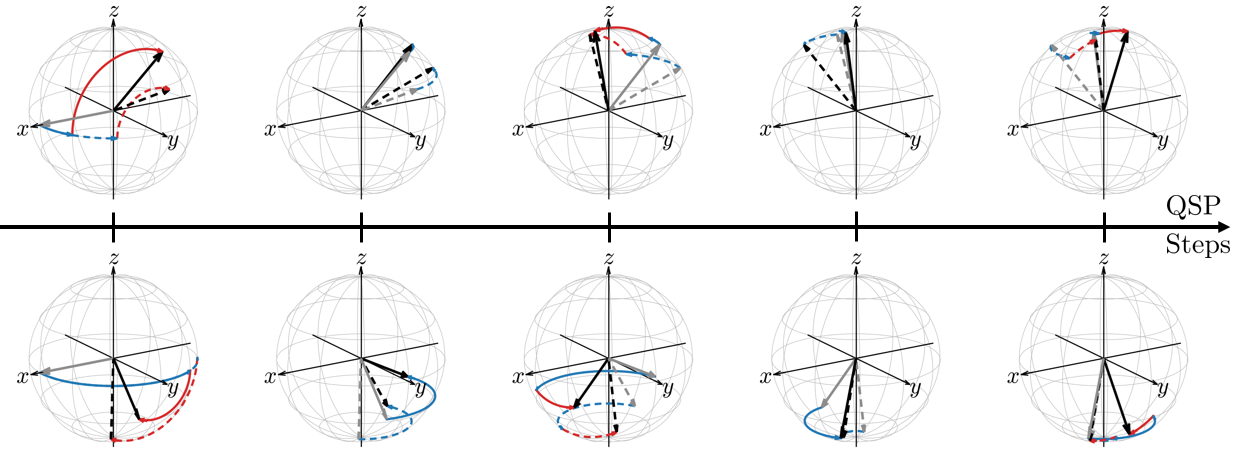}
    \caption{An example 5-step QSP sequence, with example trajectories on the Bloch sphere starting from $\ket{+}$ for $\Phi = \pi/6$ (top row, solid), $\Phi = \pi/3$ (top row, dashed), $\Phi = 2\pi/3$ (bottom row, solid), and $\Phi = 5\pi/6$ (bottom row, dashed). In each plot, the blue arrows show signal rotations; the red arrows show control rotations (which are the same in the top and bottom rows; no controls are applied in the second or fourth columns); the gray arrows show the states before the two rotations in that plot; and the black arrows show the states after those one or two rotations. All four outcomes are ``correct''---i.e. the small $\Phi$ values in the top row end up in the $\ket{1}$ state, and the large $\Phi$ values in the bottom row end up in the $\ket{0}$ state---with probability at least $97.4\%$; see Fig.~\ref{fig:QSPOutcome}.}
    \label{fig:QSPSequence}
\end{figure}

\begin{figure}[t]
    \includegraphics[width=0.45\columnwidth]{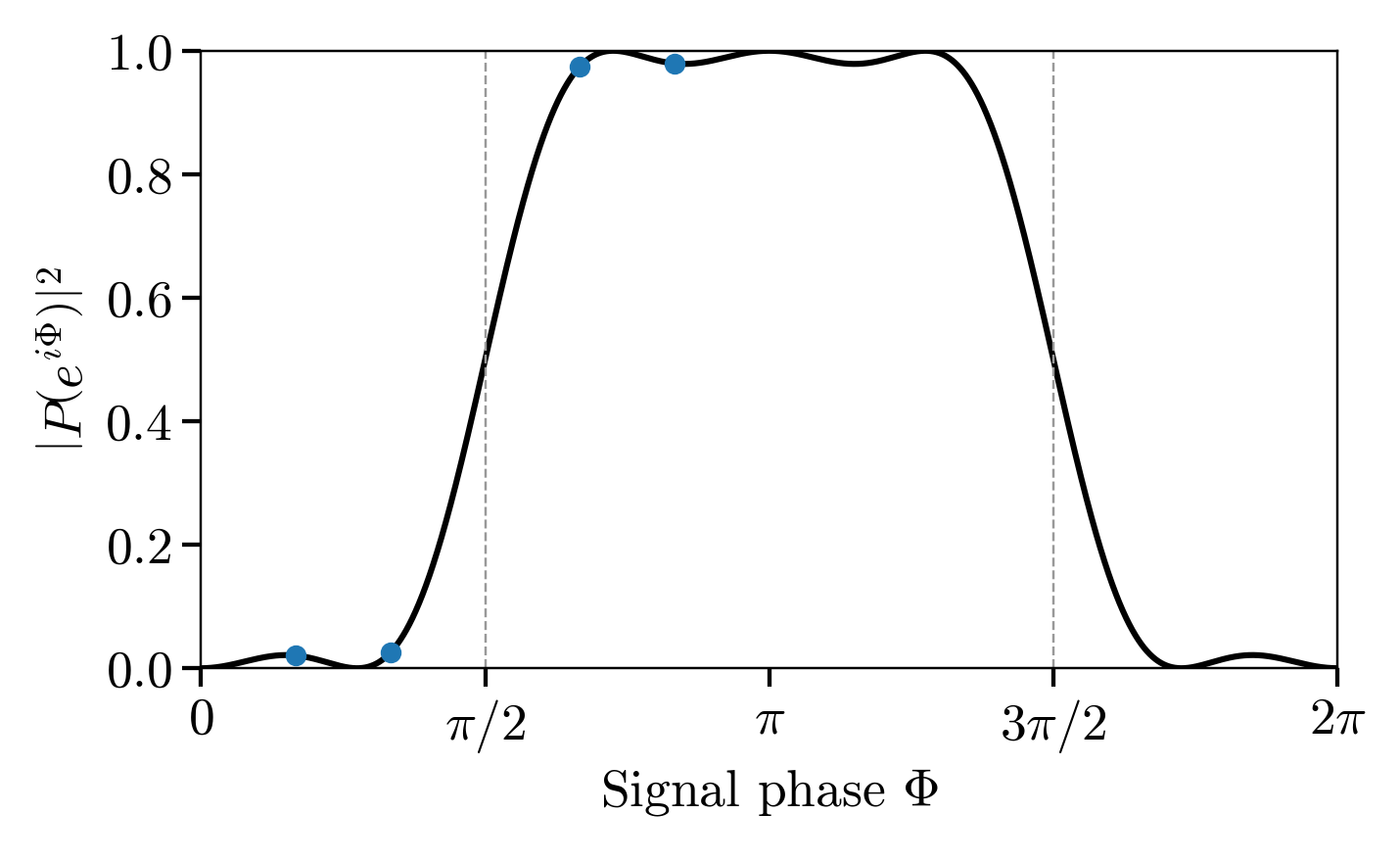}
    \caption{The value of $\left|P\left(e^{i\Phi}\right)\right|^2$, i.e. the probability of the example QSP sequence in Fig.~\ref{fig:QSPSequence} leading to a final state that is $\ket{0}$, as a function of the signal phase $\Phi$. The transitions around $\Phi = \pm \pi/2$ are shown as dashed vertical lines, while the four example trajectories in Fig.~\ref{fig:QSPSequence} correspond to the blue dots.}
    \label{fig:QSPOutcome}
\end{figure}

Our example QSP, using $N_\QSP = 5$ iterations (i.e. 5 applications of $U'_\Phi$) is illustrated in Fig.~\ref{fig:QSPSequence}. The top row of Fig.~\ref{fig:QSPSequence} illustrates the evolution, starting from the $\ket{+}$ state (i.e. beginning with a controlled rotation by $\pi/2$ about the $y$ axis, if one starts with $\ket{0}$; we omit this initial control rotation from the plots for visual clarity), for $\Phi = \pi/6$ (solid) and $\pi/3$ (dashed); the bottom row repeats this for $2\pi/3$ (solid) and $5\pi/6$ (dashed). In each plot, the blue arrows show the $\Phi$-dependent rotation induced by the signal, while the red arrows illustrate the fixed control operations; no control is applied in the second or fourth columns. The goal of the QSP is thus to reach the north pole $\ket{1}$ in the top row and the south pole $\ket{0}$ in the bottom row, which is done with high accuracy for the illustrated angles.

The performance for general $\tilde{\Phi}$ is shown in Fig.~\ref{fig:QSPOutcome}. As that plot shows, this particular QSP performs well for $|\tilde{\Phi}| < \pi/3$ and for $|\tilde{\Phi}| > 2\pi/3$, but it has a wide transition region between those regimes. More generally, a transition width $\Delta$ and probability error $\xi$ can be achieved with
\begin{align}
    N_{\rm QSP} = O\left( \frac{1}{\Delta} \ln\pfrac{1}{\xi} \right)
\end{align}
iterations.

\bibliographystyle{apsrev4-2}
\bibliography{main}

\end{document}